\documentclass{article}%
\usepackage{graphicx}
\usepackage{amsmath}
\usepackage{amsfonts}
\usepackage{amssymb}%
\usepackage{upgreek}
\usepackage{placeins}
\usepackage{subfigure}
\usepackage{float}
\usepackage{adjustbox}
\usepackage{cite}
\usepackage{hyperref}
\providecommand{\U}[1]{\protect\rule{.1in}{.1in}}

\makeatletter
  \renewcommand\thefootnote{\fnsymbol{footnote}}
\makeatother
\begin{document}

\author{Antony Valentini\\Augustus College}

\begin{center}
{\LARGE Towards a test of the Born rule in high-energy collisions}
\end{center}

\bigskip

\bigskip

\begin{center}
Antony Valentini\footnote{email: a.valentini@imperial.ac.uk}

\textit{Abdus Salam Centre for Theoretical Physics, Imperial College London,}

\textit{Prince Consort Road, London SW7 2AZ, United Kingdom.}

Mira Varma\footnote{email: mira.varma@yale.edu}

\textit{Department of Physics, Yale University, New Haven, CT 06520, USA.}
\end{center}

\makeatletter
  \renewcommand\thefootnote{\arabic{footnote}}
  \setcounter{footnote}{0} 
\makeatother
\bigskip

\bigskip

\bigskip

We consider how the Born rule, a fundamental principle of quantum mechanics,
can be tested for particles created on the shortest timescales
($\sim10^{-25}\,\mathrm{s}$) currently accessible at high-energy colliders. We
focus on targeted tests of the Born rule for spin or polarisation
probabilities, which offer a particularly clean experimental signal, and which
can be described by a simple hidden-variables model of two-state systems
proposed by Bell. These probabilities test a remarkable feature of the quantum
formalism, whereby expectation values for incompatible experiments are
linearly related. Born-rule violations can be parameterised by nonlinear
expectation values for quantum measurements of spin or polarisation, together  
with anomalies in ensemble averages, which may then be constrained by
experiment. Notable experiments considered here include the recent detection
of single photons from top-quark decay, and the indirect measurement of
tau-lepton polarisation. Repurposing these experiments as tests of the Born
rule, however, presents several challenges, which are discussed in this paper.

\bigskip

\bigskip

\bigskip

\bigskip

\bigskip

\bigskip

\bigskip

\bigskip

\bigskip

\bigskip

\bigskip

\bigskip

\bigskip

\bigskip

\bigskip

\bigskip

\bigskip

\bigskip

\bigskip

\section{Introduction}
\label{sec:introduction}

The domain of validity of any scientific theory can be determined only by
experiment. It is therefore important that we continue to test our most
fundamental physical theories -- including quantum mechanics -- in extreme and
novel conditions. For several decades much effort was devoted to testing the
peculiarities of quantum entangled systems, confirming in particular the
presence of nonlocality as evidenced by violations of Bell's
inequality.\footnote{Up to some loopholes~\cite{MyrStanf}.} Such tests were
generally carried out at low energies, for example with optical photons. In
recent years, similar tests have been extended into the high-energy domain, by
repurposing collision data to obtain evidence for entanglement and nonlocality
at energy scales far beyond those previously studied~\cite{Barretal}.
Similarly, tests of the linearity of the Schr\"{o}dinger equation were carried
out at low energies~\cite{Wein,Winel}, and it would be of interest to extend
such tests into the high-energy domain where nonlinearities can generate new
effects~\cite{Kaplan}. In this paper we consider another fundamental principle of quantum mechanics -- the Born probability rule -- exploring how it might be tested in the high-energy domain and at the shortest timescales ($\sim 10^{-25}\,\mathrm{s}$) currently observable at colliders. While the statistical results of any quantum experiment depend on the Born rule, of interest here are targeted tests that probe the Born rule as far as possible to high precision and in extreme domains.

As a matter of principle, it has been argued that the Born rule may be less fundamental than the other axioms of quantum mechanics, prompting several attempts to derive or justify it in some way~\cite{DD99,DWb,Fuchs,SS,Zurek}.
It has also been suggested that quantum physics with the Born rule may be
merely a special case of a wider physics in which the Born rule is broken.
This is in fact a natural implication of (nonlocal) hidden-variables theories,
such as de Broglie-Bohm pilot-wave theory~\cite{deB28,BV09,B52a,B52b,Holl93,AVEMP24,AV26}, in which the Born rule arises
as a state of statistical equilibrium (analogous to thermal equilibrium in
classical physics)~\cite{AV91a,AV91b,AV92,VW05,Allori}: more general `quantum
nonequilibrium' ensembles, with Born-rule violations, are allowed in principle~\cite{AV91a,AV91b,AV92,AV02Pram,AV02c,AVEMP24,AVOldo,AVPop25,AV26}. Another
approach, based on `quantum measure theory', considers how the Born rule might
be violated by a set of probabilistic axioms more general than those usually
employed in quantum theory~\cite{Sorkin}.\footnote{See Ref.~\cite{Kite} for a
comparative study of de Broglie-Bohm theory and quantum measure theory.}

Experimentally, quantum nonequilibrium violations of the Born rule have been
searched for in data from the cosmic microwave background (CMB)~\cite{AV10,VPV}. According to inflationary cosmology, temperature anisotropies
in the CMB were ultimately generated by Born-rule quantum fluctuations in a
primordial inflaton field in the very early universe~\cite{LL00,PUz}.
Precision data from the CMB then allows us to put constraints on violations of
the Born rule at very early times~\cite{AV07,AV10,CV13,CV15,VPV}. Low-energy
laboratory tests of the Born rule have also been carried out, searching in
particular for the novel interference effects suggested by generalised quantum
measure theory~\cite{Laf10}.

More recently it has been argued that, in the de Broglie-Bohm formulation of
canonical quantum gravity, the Born rule may be gravitationally unstable at
the Planck scale~\cite{AV21,AV23}. Such effects could in principle manifest
today as a small violation of the Born rule in Hawking radiation from
exploding primordial black holes, whose gamma-ray emissions are being searched
for~\cite{Ack18}. For such rays, however, tests of the Born rule would require
a dedicated satellite in space to avoid absorption by the earth's atmosphere.
This and related suggestions have motivated a test of the Born rule aboard the recently launched QUICK%
${{}^3}$
satellite mission, which includes an interferometer designed to test the Born
rule for photons from an onboard optical source~\cite{Ahm24}. Once the
proof-of-concept experiment is fully operational, in the future it could be
redeployed to test the Born rule for photons from exotic astrophysical sources, such as exploding primordial black holes~\cite{AV21,AV23} or disintegrating dark matter~\cite{UV20}.

In this paper we consider how high-energy collision experiments might be
repurposed as tests of the Born rule at the shortest accessible timescales.
Two approaches have been suggested~\cite{AnnFond}: (i) to test the Born rule
via measurements of high-energy differential scattering cross sections, and
(ii) to test the Born rule via measurements of spin or polarisation
probabilities for particles emerging from high-energy collisions.

In the first approach, applying the Born rule to calculate the probability
$\left\vert \left\langle f\right\vert \hat{S}\left\vert i\right\rangle
\right\vert ^{2}$ for a transition $i\rightarrow f$ from an initial state
$\left\vert i\right\rangle $ to a final state $\left\vert f\right\rangle $
(where $\left\langle f\right\vert \hat{S}\left\vert i\right\rangle $ is the
associated $S$-matrix element), we obtain the usual result%
\begin{equation}
\left(  \frac{d\sigma}{d\Omega}\right)  _{\mathrm{QT}}\propto\left\vert
\mathcal{M}\right\vert ^{2}\,,\label{dsigma}%
\end{equation}
whereby the quantum-theoretical differential scattering cross section
$(d\sigma/d\Omega)_{\mathrm{QT}}$ for the process is proportional to the
modulus-squared of the Feynman amplitude $\mathcal{M}$ (obtained from the
associated Feynman diagrams). In a sense, any test of the standard prediction
(\ref{dsigma}) is in effect a test of the Born rule. However, current
experiments measuring cross sections at colliders do not perform targeted
tests, aimed specifically at setting precise limits on possible Born-rule violations.

It has been suggested that, in collision experiments, kinematical regions
where $\left\vert \mathcal{M}\right\vert ^{2}$ is predicted to be very small,
or even zero, might provide especially sensitive probes of the Born rule~\cite{AnnFond}. Certainly, general constraints on Born-rule violations can be
obtained by matching simple deformations of (\ref{dsigma}) with collision
data: for example, we might take%
\begin{equation}
\frac{d\sigma}{d\Omega}\propto\left\vert \mathcal{M}\right\vert
^{2+\varepsilon}\,,
\end{equation}
with $\varepsilon$ a small parameter, or simply smear the density $\left\vert
\mathcal{M}\right\vert ^{2}$ with a narrow Gaussian in the relevant
kinematical space~\cite{VVprep}. However, instead of signalling a violation of
the Born rule, observed anomalies in scattering cross sections might be
attributed to the presence of an unknown particle (contributing to Feynman
diagrams), or to some other unexpected correction to the underlying
Lagrangian, or possibly to some ambiguity or error in the computation of
higher-order corrections. For this reason, in this paper we focus on the
second approach, which appears, at least initially, to provide a cleaner test
of the Born rule via the measurement of spin or polarisation probabilities for
particles emerging from high-energy collisions. As we will see, however, there
are challenges.

Testing the Born rule via spin or polarisation probabilities also provides a
direct test of a general and remarkable feature of quantum mechanics, whereby
expectation values are additive across physically incompatible experiments
(performed to measure distinct non-commuting quantum observables, in this case
spin or polarisation along different axes in space)~\cite{AV04}. While this
peculiar feature has not received as much attention as quantum nonlocality, or
quantum contextuality, as Bell rightly emphasised it is `a quite peculiar
property of quantum mechanical states'~\cite{Bell66}.\footnote{See also Refs.~\cite{Mermin93,Timpson04}.} It was argued long ago that a breakdown of additive
expectation values would signal the presence of quantum nonequilibrium in a
general class of nonlocal hidden-variables theories, and that such tests might
be performed for photons from exotic astrophysical sources~\cite{AV04}. In
this paper, we argue that such tests can and should be performed for photons
and other particles created by short-timescale high-energy collisions.

To account for quantum measurements of spin or polarisation, it suffices to
consider the Bell model of two-state systems~\cite{Bell66,Mermin93}. This is a
simple example of a deterministic hidden-variables theory, in which
measurement outcomes for an individual system are determined by a hidden unit
vector. The ensemble Born rule emerges only when the hidden variables have a
specific equilibrium distribution (uniform on the Bloch sphere). For more
general distributions, the Born rule is broken. Furthermore, expectation
values become nonlinear or non-additive (breaking a basic principle of quantum
mechanics), and ensemble averages can be anomalous. These effects provide
simple parameterisations of Born-rule violations, to be constrained by
experiment.\footnote{Pilot-wave theory can be extended to a model in which
violations of the Born rule are generated by high-energy collisions~\cite{AnnFond}, providing an example of the phenomenology considered in this
paper. However, Bell's simple model of two-state systems suffices for our
purposes.}

As a simple example, if photons pass through a linear polariser, followed by a
second linear polariser at angle $\Theta$ with respect to the first, quantum
theory unambiguously predicts that the transmission probability
$p_{\mathrm{QT}}^{+}$ through the second polariser will vary with $\Theta$
according to Malus' law:%
\begin{equation}
p_{\mathrm{QT}}^{+}(\Theta)=\cos^{2}\Theta\,.\label{Malus}%
\end{equation}
Should deviations from (\ref{Malus}) be observed, for single photons
transmitted one at a time through a pair of polarisers, it would provide a
strong signal that the Born rule was broken.

A similar experiment was in fact carried out in 1967 by Papaliolios~\cite{Pap67}, with optical photons from a tungsten filament lamp. The purpose
was to test a hidden-variables model due to Bohm and Bub~\cite{BB66} (distinct
from de Broglie-Bohm theory), in which deviations from the Born rule are
predicted to exist for a very short time (perhaps of order $10^{-13}%
\,\mathrm{s}$) immediately after a quantum measurement has taken place.
According to the model, if a second quantum measurement takes place within
(say) $\sim10^{-13}\,\mathrm{s}$ of the first, deviations from the Born rule
should be observed. To test this, Papaliolios placed three linear polarisers
in close proximity and varied the angle of the third. The first polariser
prepared a linearly-polarised quantum state, the second measured the
polarisation along an axis almost orthogonal to the first, while the third
performed a second measurement at an angle $\Theta$ relative to the first
measurement. By placing the polarisers sufficiently close together to
guarantee the short timescale, Papaliolios verified that the transmission
probability through the third polariser varied with $\Theta$ according to
(\ref{Malus}) within an accuracy of about $1\%$, thereby confirming the
approximate validity of the Born rule over timescales of order $10^{-13}%
\,\mathrm{s}$.\footnote{Historically, this appears to have been the first
experiment designed to test a hidden-variables theory~\cite{Freire}.}

In this paper we explore how targeted tests of the Born rule, for spin or
polarisation probabilities, might be carried out at much higher energies, and
for particles created over timescales of order $10^{-25}\,\mathrm{s}$ (about
12 orders of magnitude smaller than the timescales probed by Papaliolios). As
we shall see, however, such tests present certain challenges.

A key challenge concerns the measurement of polarisation in the high-energy regime. In this paper, we will be considering polarisation measurements in particular for gamma-ray photons. To our knowledge, there appears to be no current technology able to measure polarisation accurately for individual photons in this energy range. There are, however, well-developed methods for measuring the mean polarisation of beams in this regime, developed mainly for use in astrophysics. For our purposes, while the measurement of individual polarisation would be ideal, measurements of mean polarisation would also be of interest. In this paper we will assume that, in the near future, polarimeters will be available for use in high-energy collision experiments that are sensitive at least to the mean polarisation in the gamma-ray region and perhaps eventually even enable the measurement of individual photon polarisation. 

For completeness, we briefly summarise the state-of-the-art in gamma-ray polarimetry. The widely used technique of Compton polarimetry exploits the fact that, in Compton scattering, photons are scattered preferentially perpendicular to the incident mean polarisation vector, a technique that works well in the energy range up to about 100 keV~\cite{DelMonte:2023irw}. Recent progress in gamma-ray polarimetry is exemplified by upcoming instruments such as POLAR-2~\cite{Produit:2023dei}, which will use Compton scattering in the soft gamma-ray region to extract mean polarisation from astrophysical gamma-ray bursts, by analysing the distribution of azimuthal scattering angles over many events. Proposed missions, including AMEGO, AMEGO-X, and e-ASTROGAM~\cite{AMEGO:2019gny,Fleischhack:2021mhc,e-ASTROGAM:2017pxr} aim to measure the mean polarisation of gamma rays in the MeV-GeV range by combining Compton scattering with electron–positron pair production. However, while all three proposed missions incorporate the necessary instrumentation to study polarisation across this energy range, their projected polarimetric sensitivity remains limited to the MeV regime, where Compton-based methods remain the most effective. It may be hoped that, in the near future, similar techniques might be applied to the study of polarisation for gamma rays emerging from high-energy collisions. 

In Section~\ref{sec:born_rule_for_two_state_systems}, we discuss the Born rule for two-state quantum systems. We outline the formalism of hidden-variables theories and show how they imply an extended physics with Born-rule violations parameterised by anomalous expectation values. These ideas are illustrated by the simple Bell model of two-state systems. In Section~\ref{sec:photons}, we discuss the detection of photons emerging from short-timescale collider processes, in particular single photons from top-quark decay and photon pairs from neutral pion decay. We consider how such experiments might be adapted to test the Born rule for photon polarisation probabilities. Section~\ref{sec:tau_leptons} focuses on average polarisation measurements for tau leptons from $Z$-boson decay and considers how these may be used to study possible deviations from the Born rule. In Section~\ref{sec:bornrule_background}, we highlight how collider experiments employ the Born rule for background modelling and event selection, and we discuss the implications this may have for our proposed tests. Finally, in Section~\ref{sec:conclusion}, we summarise our findings and discuss the broader significance of possible Born-rule violations at high energies.

\section{The Born rule for two-state systems}
\label{sec:born_rule_for_two_state_systems}

Our focus will be on photon polarisation, as well as on lepton spin. Hence it
suffices to consider the formalism for a general two-state quantum system,
with quantum observables $\hat{\sigma}\equiv\mathbf{m}\cdot\boldsymbol{\hat
{\upsigma}}$ and eigenvalues $\sigma=\pm1$, where $\mathbf{m}$ is a unit vector
on the Bloch sphere and $\boldsymbol{\hat{\upsigma}}$ is the Pauli spin
operator. If we measure $\hat{\sigma}$ over an ensemble of systems with
density operator $\hat{\rho}$, quantum theory predicts an expectation value%
\begin{equation}
E_{\mathrm{QT}}(\mathbf{m})\equiv\left\langle \mathbf{m}\cdot\boldsymbol{\hat
{\upsigma}}\right\rangle =\mathrm{Tr}\left[  \hat{\rho}\left(  \mathbf{m}%
\cdot\boldsymbol{\hat{\upsigma}}\right)  \right]  =\mathbf{m}\cdot
\mathbf{P}_{\mathrm{QT}}\mathbf{\ ,} \label{E_QT}%
\end{equation}
where%
\begin{equation}
\mathbf{P}_{\mathrm{QT}}=\langle\boldsymbol{\hat{\upsigma}}\rangle
=\mathrm{Tr}\left[  \hat{\rho}\boldsymbol{\hat{\upsigma}}\right]  \label{PQT}%
\end{equation}
is the quantum-theoretical average polarisation for the ensemble. For a
measurement of an individual system, the outcome $\sigma=+1$ has a Born-rule
probability%
\begin{equation}
p_{\mathrm{QT}}^{+}(\mathbf{m})=\frac{1}{2}\left(  1+E_{\mathrm{QT}%
}(\mathbf{m})\right)  =\frac{1}{2}\left(  1+P_{\mathrm{QT}}\cos\theta\right)
\ , \label{Malus_gen}%
\end{equation}
where $\theta$ is the angle between $\mathbf{m}$ and $\mathbf{P}_{\mathrm{QT}%
}$. Deviations from the sinusoidal modulation (\ref{Malus_gen}) would signal a
violation of the usual quantum Born rule.
 
For particles of spin $\tfrac{1}{2}$, $\sigma=\pm1$ indicates spin (in units $\hbar/2$)
up or down a spatial axis $\mathbf{m}$. For photons, $\sigma=\pm1$ indicates
polarisation parallel or perpendicular to a spatial axis $\mathbf{M}$, where
now $\theta$ corresponds to an angle $\Theta=\theta/2$ in 3-space.

For single photons passing through a linear polariser, the emerging beam is
fully polarised ($P_{\mathrm{QT}}=1$). If the beam is then directed towards a
second linear polariser, at angle $\Theta$ with respect to the first, the
general result (\ref{Malus_gen}) predicts the transmission probability
(\ref{Malus}) tested by Papaliolios. Again, deviations from (\ref{Malus})
would signal a breakdown of the Born rule.

\begin{figure}[H]
\centering
\adjustbox{trim={0 0cm 0 0cm},clip}{
\includegraphics[width=0.55\textwidth]{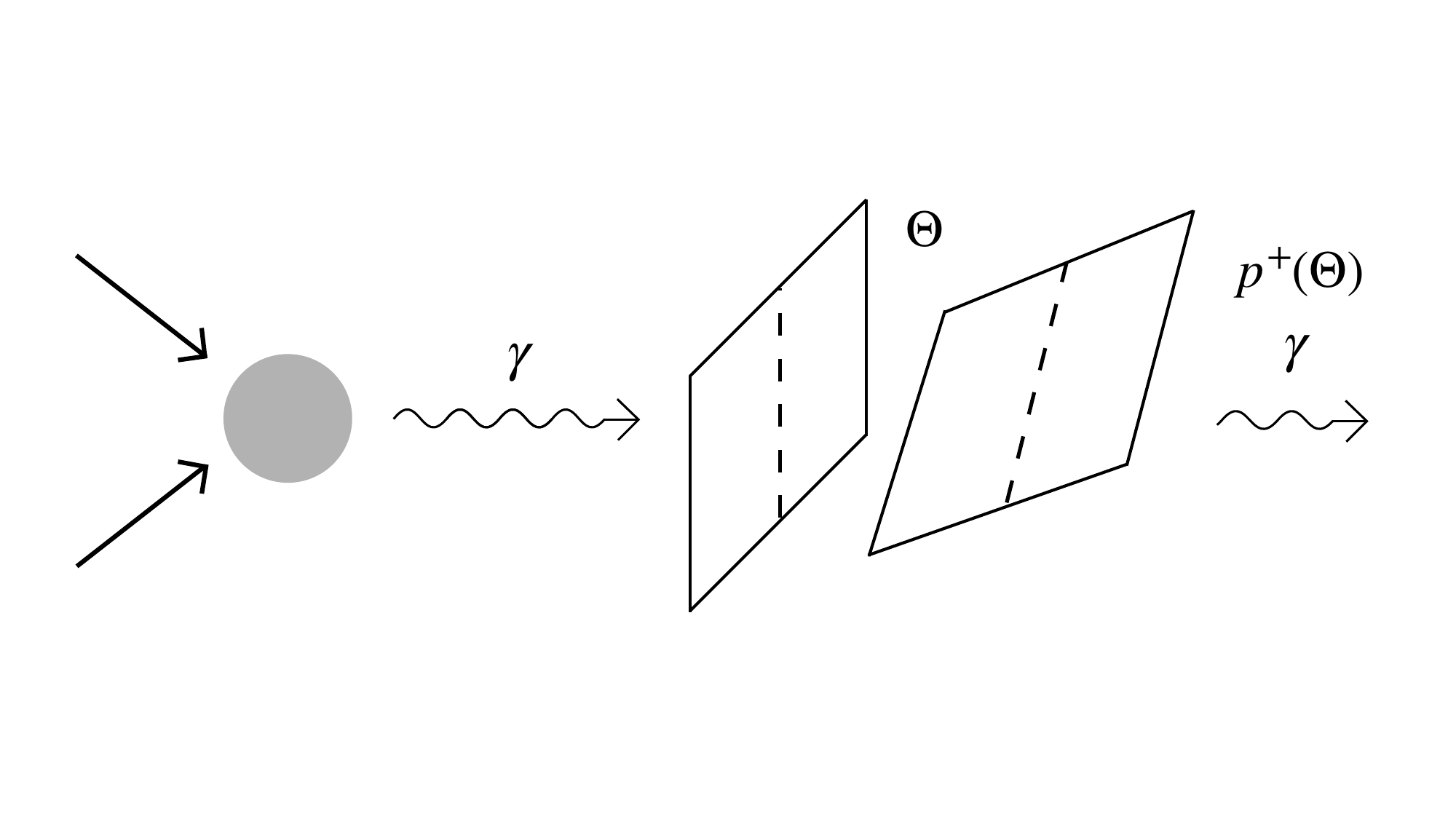}
}
\caption{An ideal test of (\ref{Malus}) for single photons emerging from a
high-energy collision.}
\label{fig:Malus_law}
\end{figure}

Ideally, we might test (\ref{Malus}) for single photons emerging from
high-energy collisions, as shown in the thought experiment of Fig.~\ref{fig:Malus_law}. Such a
simple scenario is, however, not practical at present, as we shall discuss.%

\subsection{Hidden variables and the Born rule}

A deterministic hidden-variables theory posits unknown parameters $\lambda$
that determine the outcomes of quantum measurements. Formally, for an
experiment specified by apparatus settings $M$, such a theory provides a
mapping $\omega=\omega(M,\lambda)$ from initial values of $\lambda$ to final
outcomes $\omega$ (equal to eigenvalues of the operator observable
$\hat{\omega}$ that is being measured). Over an ensemble of experiments with
fixed $M$, the hidden variables $\lambda$ will generally vary from one run of
the experiment to the next. For a general ensemble distribution $\rho
(\lambda)$, the expectation value of $\omega$ will be\footnote{Here $\int
d\lambda$ is understood as a generalised sum, in case $\lambda$ is not
continuous.}%
\begin{equation}
\left\langle \omega\right\rangle =\int d\lambda\ \rho(\lambda)\omega
(M,\lambda)\ .
\end{equation}
In a properly constructed theory, there is a particular distribution
$\rho_{\mathrm{QT}}(\lambda)$ for which the corresponding expectation value
$\left\langle \omega\right\rangle _{\mathrm{QT}}$ matches the usual Born-rule:%
\begin{equation}
\left\langle \omega\right\rangle _{\mathrm{QT}}=\mathrm{Tr}(\hat{\rho}%
\hat{\omega})
\end{equation}
(for some density operator $\hat{\rho}$ ). It is, however, important to note
that if $\rho(\lambda)\neq\rho_{\mathrm{QT}}(\lambda)$, then in general we
will find%
\begin{equation}
\left\langle \omega\right\rangle \neq\left\langle \omega\right\rangle
_{\mathrm{QT}}%
\end{equation}
and the Born rule will be broken. In this way, the theory can account for
quantum physics with the assumed distribution $\rho_{\mathrm{QT}}(\lambda)$.
But at least in principle, the theory also implies an extended physics beyond
the Born rule, with more general distributions $\rho(\lambda)\neq
\rho_{\mathrm{QT}}(\lambda)$~\cite{AV02c,AV04,AV07}.

Pilot-wave theory provides a concrete example~\cite{deB28,BV09,B52a,B52b,Holl93,AVEMP24,AV26}. The outcome $\omega$ of a
quantum measurement is determined by the initial (hidden) configuration $q(0)$
of the system, together with the initial wave function $\psi(q,0)$, so that
$\lambda=(q(0),\psi(q,0))$. For an ensemble of experiments with the same
$\psi(q,0)$, and the same apparatus settings $M$, in effect we have
$\lambda=q(0)$ and the quantum-theoretical distribution $\rho_{\mathrm{QT}%
}(\lambda)$ is simply given by the Born rule $\rho_{\mathrm{QT}}%
(q,0)=\left\vert \psi(q,0)\right\vert ^{2}$ in configuration space. For more
general `quantum nonequilibrium' distributions $\rho(q,0)\neq\left\vert
\psi(q,0)\right\vert ^{2}$ we obtain a more general physics with violations of
the Born rule~\cite{AV91a,AV91b,AV92,AV02Pram,AVEMP24,AVOldo,AVPop25,AV26},
which as noted has been extensively discussed in particular in a cosmological
context~\cite{AV07,AV10,CV13,CV15,VPV}.

For the simple case of a two-state system, with quantum observables
$\hat{\sigma}=\mathbf{m}\cdot\boldsymbol{\hat{\upsigma}}$, a deterministic
hidden-variables theory provides a mapping $\sigma=\sigma\left(
\mathbf{m},\lambda\right)  $ that determines the outcomes $\sigma=\pm1$ for a
given measurement axis $\mathbf{m}$. For a particular ensemble distribution
$\rho_{\mathrm{QT}}(\lambda)$, we find the Born-rule expectation value%
\begin{equation}
E_{\mathrm{QT}}(\mathbf{m})\equiv\left\langle \sigma\left(  \mathbf{m}%
,\lambda\right)  \right\rangle _{\mathrm{QT}}=\int d\lambda\ \rho
_{\mathrm{QT}}(\lambda)\sigma\left(  \mathbf{m},\lambda\right)  =\mathbf{m}%
\cdot\mathbf{P}_{\mathrm{QT}}\,.\label{E_QT2}%
\end{equation}
However, for $\rho(\lambda)\neq\rho_{\mathrm{QT}}(\lambda)$, in general the
expectation value%
\[
E(\mathbf{m})\equiv\left\langle \sigma\left(  \mathbf{m},\lambda\right)
\right\rangle =\int d\lambda\ \rho(\lambda)\sigma\left(  \mathbf{m}%
,\lambda\right)
\]
will be unequal to $E_{\mathrm{QT}}(\mathbf{m})$, and moreover will not take
the linear form $\mathbf{m}\cdot\mathbf{P}$ for any fixed Bloch vector
$\mathbf{P}$. The outcome probability%
\begin{equation}
p^{+}(\mathbf{m})=\frac{1}{2}\left(  1+E(\mathbf{m})\right)  \label{p+}%
\end{equation}
will then disagree with the Born-rule result (\ref{Malus_gen})~\cite{AV04}.

As noted in the Introduction, deviations from the Born rule are also signalled
by a breakdown of additive expectation values for non-commuting observables.
As is easily demonstrated, the linearity in $\mathbf{m}$ of the quantum
expectation value (\ref{E_QT2}) is equivalent to additive expectation values 
for arbitrary incompatible observables $\hat{\sigma}=\mathbf{m}\cdot
\boldsymbol{\hat{\upsigma}}$ and $\hat{\sigma}^{\prime}=\mathbf{m}^{\prime}%
\cdot\boldsymbol{\hat{\upsigma}}$ (with $\mathbf{m}^{\prime}\neq\mathbf{m}$)~\cite{AV04}.

\subsection{Parameterising deviations from the Born rule}

We can parameterise anomalous (non-Born-rule) probabilities $p^{+}%
(\mathbf{m})$ by a spherical harmonic expansion%
\begin{equation}
p^{+}(\theta,\phi)=\sum_{l=0}^{\infty}\sum_{m=-l}^{+l}b_{lm}Y_{lm}(\theta
,\phi)\ ,\label{p+sphhar}%
\end{equation}
with $\mathbf{m}$ specified by angular coordinates $(\theta,\phi)$ on the
Bloch sphere and $\mathbf{P}_{\mathrm{QT}}$ pointing along $+z$.\footnote{The
reality of $p^{+}$ implies $b_{lm}^{\ast}=(-1)^{m}b_{l(-m)}$.} In quantum
theory, $p^{+}=p_{\mathrm{QT}}^{+}$ takes the form (\ref{Malus_gen}) with a
simple dipole term in $\mathbf{m}$, and the only non-zero coefficients are%
\begin{equation}
(b_{00})_{\mathrm{QT}}=\sqrt{\pi}\ ,\ (b_{10})_{\mathrm{QT}}=P_{\mathrm{QT}}\sqrt{\pi/3}\ .
\end{equation}
In an extended theory, the probability $p^{+}$ could receive higher multipole corrections.

We may also parameterise the deviations in terms of anomalous expectation
values~\cite{AV07},%
\begin{equation}
E(\mathbf{m})=\epsilon+P_{i}m_{i}+Q_{ij}m_{i}m_{j}+R_{ijk}m_{i}m_{j}%
m_{k}+...\label{E_anom}%
\end{equation}
(summing over repeated indices), where in Bloch space $\epsilon$ is a constant
scalar, $P_{i}$ is a constant vector, and $Q_{ij}$, $R_{ijk}$, ... are
constant tensors, and where in general $P_{i}\neq(P_{i})_{\mathrm{QT}}$.
Anomalous outcome probabilities (\ref{p+}) may then be written directly in
terms of the parameters $\epsilon$, $P_{i}$, $Q_{ij}$, $R_{ijk}$, ... .

By carefully monitoring spin or polarisation probabilities for two-state
systems, we may search for a possible breakdown of expectation linearity and
of the Born rule. Known experiments appear to be consistent with the
quantum-theoretical values%
\begin{equation}
\epsilon=Q_{ij}=R_{ijk}=...=0\,,\,\,P_{i}=(P_{i})_{\mathrm{QT}}\,.
\end{equation}
The experimental challenge addressed in this paper is how best to set limits
on deviations from these values, at the shortest timescales currently
accessible at colliders.

\subsection{The Bell model}
\label{subsec:the_bell_model}
We can illustrate the above general reasoning by means of a simple
hidden-variables model of two-state systems proposed by Bell~\cite{Bell66,Mermin93}.

Bell formulated his model for a pure quantum state $\left\vert \psi
\right\rangle $, for which $\mathbf{P}_{\mathrm{QT}}$ has unit norm
($P_{\mathrm{QT}}=1$). It may be extended to a mixed state by applying the
original model to appropriate pure sub-ensembles.

In Bell's pure-state model, each individual system has a hidden unit vector
$\boldsymbol{\lambda}$ which determines the outcome $\sigma=\pm1$ (of a
quantum measurement of $\hat{\sigma}=\mathbf{m}\cdot\mathbf{\hat{\sigma}}$) by
means of the mapping\footnote{Cases where $(\boldsymbol{\lambda}%
+\mathbf{P}^{\mathrm{QT}})\cdot\mathbf{m}=0$ form a set of measure zero and
may be ignored, though of course an outcome could be specified for these as
well.}%
\begin{equation}
\sigma = \sigma\!\left(\mathbf m,\boldsymbol\lambda\right)=
\left\{
\begin{array}{ll}
  +1, & \text{if } (\boldsymbol\lambda+\mathbf P_{\mathrm{QT}})\!\cdot\!\mathbf m > 0,\\
  -1, & \text{if } (\boldsymbol\lambda+\mathbf P_{\mathrm{QT}})\!\cdot\!\mathbf m < 0 .
\end{array}
\right .\label{Bell map}%
\end{equation}
This is illustrated in Fig. 2, 
where we show the unit sphere in Bloch space, together with the unit vectors $\mathbf{m}$, $\mathbf{P}%
_{\mathrm{QT}}$ and $\boldsymbol{\lambda}$. The shaded region indicates the surface area $A_{-}$ of the sphere with
$\boldsymbol{\lambda}$ such that $(\boldsymbol{\lambda}+\mathbf{P}%
_{\mathrm{QT}})\cdot\mathbf{m}<0$, for which quantum measurements yield
outcomes $\sigma=-1$. The remaining surface area $A_{+}$ yields outcomes
$\sigma=+1$.
\begin{figure}[H]
\centering
\adjustbox{trim={0 0cm 0 0cm},clip}{
\includegraphics[width=0.55\textwidth]{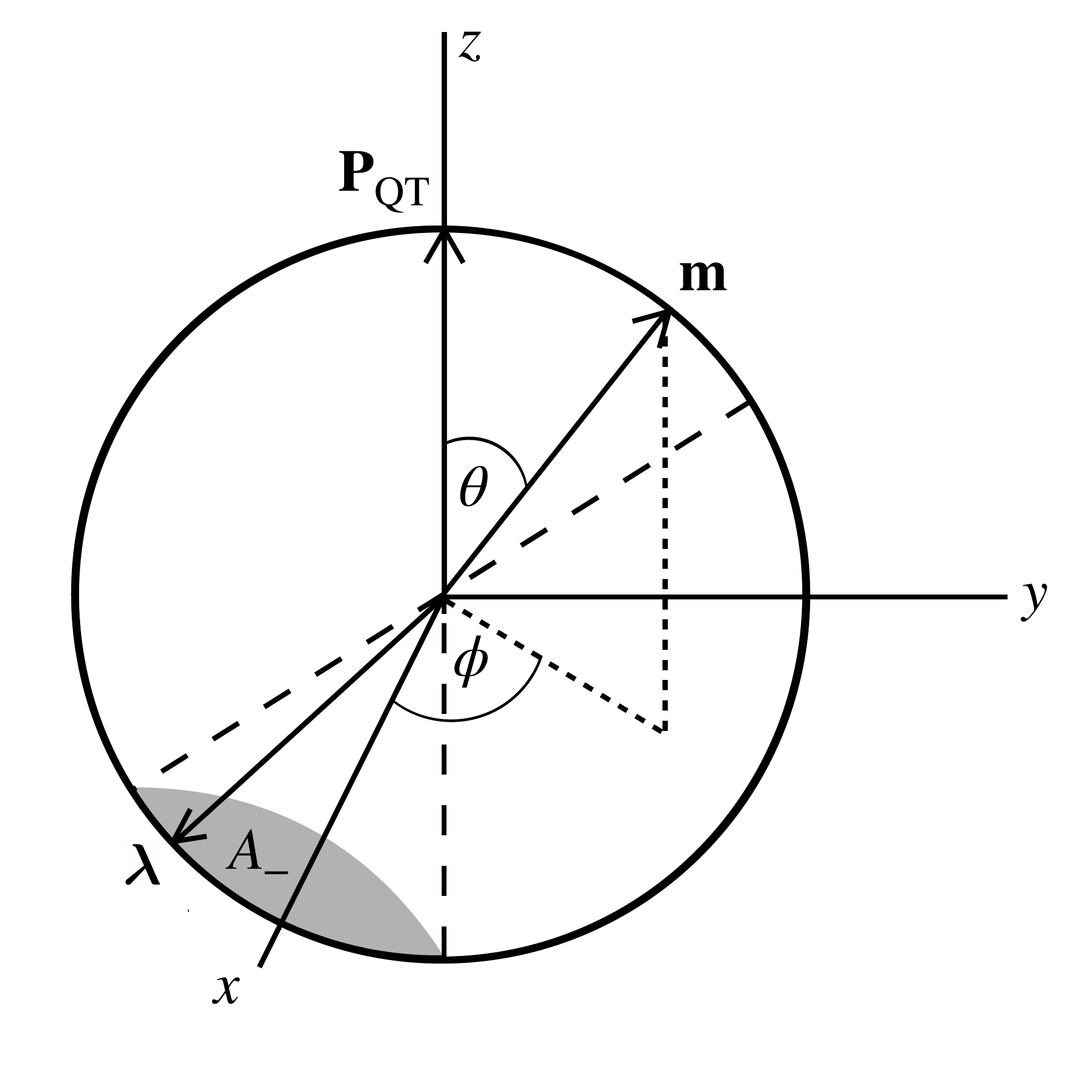}
} 
\caption{Bell's hidden-variables model of a two-state quantum system. In the
example shown, $(\boldsymbol{\lambda}+\mathbf{P}_{\mathrm{QT}})\cdot \mathbf{m}<0$ and the outcome of the quantum measurement is $\sigma=-1$.}
\label{fig:Bell_model}
\end{figure} 

For an ensemble of systems with the same quantum state, the hidden vector
$\boldsymbol{\lambda}$ generally varies from one system to another. To
reproduce the Born rule for quantum measurement outcomes, the vectors
$\boldsymbol{\lambda}$ must be distributed uniformly on the surface of the
unit sphere. Defining probability density with respect to the area element
$d\Omega$, the `equilibrium' distribution%
\begin{equation}
\rho_{\mathrm{QT}}(\boldsymbol{\lambda})=\frac{1}{4\pi}%
\end{equation}
is readily shown to yield the Born-rule expectation value%
\begin{equation}
E_{\mathrm{QT}}(\mathbf{m})=\int d\Omega\ \rho_{\mathrm{QT}}%
(\boldsymbol{\lambda})\sigma\left(  \mathbf{m},\boldsymbol{\lambda}\right)
=\mathbf{m}\cdot\mathbf{P}_{\mathrm{QT}}=\cos\theta
\end{equation}
for the outcomes $\sigma=\pm1$. To see this geometrically, note that%
\begin{equation}
p_{\mathrm{QT}}^{-}(\mathbf{m})=\int_{A_{-}}d\Omega\ \rho_{\mathrm{QT}%
}(\boldsymbol{\lambda})=\frac{1}{4\pi}A_{-}\,,
\end{equation}
where $A_{-}$ is the surface area of the sphere with $(\boldsymbol{\lambda
}+\mathbf{P}_{\mathrm{QT}})\cdot\mathbf{m}<0$ (shown shaded in Fig.~\ref{fig:Bell_model}).
Similarly,%
\[
p_{\mathrm{QT}}^{+}(\mathbf{m})=\int_{A_{+}}d\Omega\ \rho_{\mathrm{QT}%
}(\boldsymbol{\lambda})=\frac{1}{4\pi}A_{+}\,,
\]
where $A_{+}$ is the surface area of the sphere with $(\boldsymbol{\lambda
}+\mathbf{P}_{\mathrm{QT}})\cdot\mathbf{m}>0$. It is straightforward to show
that%
\begin{equation}
A_{\pm}=2\pi(1\pm\cos\theta)\label{Areas}%
\end{equation}
(where of course $A_{-}+A_{+}=4\pi$). We then indeed find an expectation value%
\begin{equation}
E_{\mathrm{QT}}(\mathbf{m})=p_{\mathrm{QT}}^{+}(\mathbf{m})-p_{\mathrm{QT}%
}^{-}(\mathbf{m})=(A_{+}-A_{-})/4\pi=\cos\theta\ .\label{EQT_Bell}%
\end{equation}

If instead $\rho(\boldsymbol{\lambda})\neq\rho_{\mathrm{QT}}%
(\boldsymbol{\lambda})$, corresponding to a non-uniform distribution on the
surface of the unit sphere, in general the outcome probabilities%
\begin{equation}
p^{\pm}(\mathbf{m})=\int_{A_{\pm}}d\Omega\ \rho(\boldsymbol{\lambda
})\label{pnoneq}%
\end{equation}
will disagree with their Born-rule values, $p^{\pm}(\mathbf{m})\neq
p_{\mathrm{QT}}^{\pm}(\mathbf{m})$, and the expectation value%
\begin{equation}
E(\mathbf{m})=p^{+}(\mathbf{m})-p^{-}(\mathbf{m})=\int d\Omega\ \rho
(\boldsymbol{\lambda})\sigma\left(  \mathbf{m},\boldsymbol{\lambda}\right)
\label{E_Bell}%
\end{equation}
will not take the simple sinusoidal form (\ref{EQT_Bell}). For a given
anomalous distribution $\rho(\boldsymbol{\lambda})$ of hidden variables, the
coefficients $\epsilon$, $P_{i}$, $Q_{ij}$, $R_{ijk}$, ... can be calculated
by performing the integrations (\ref{pnoneq})~\cite{Vprep}.

Inspection of (\ref{Bell map}) reveals the symmetry $\sigma\left(
-\mathbf{m},\boldsymbol{\lambda}\right)  =-\sigma\left(  \mathbf{m}%
,\boldsymbol{\lambda}\right)  $ of the Bell model.\footnote{For a
hidden-variables model of spin measurements that breaks this symmetry, see Ref.~\cite{Cracow}.} The general expectation value (\ref{E_Bell}) then always
satisfies $E(-\mathbf{m})=-E(\mathbf{m})$. Thus, in the Bell model, for the
anomalous expectation value (\ref{E_anom}) we might expect to obtain only odd terms in
$\mathbf{m}$:\footnote{Though a pseudoscalar term $\epsilon$, with $\epsilon \rightarrow -\epsilon$ under $\mathbf{m} \rightarrow -\mathbf{m}$, is also possible~\cite{Vprep}.}%
\begin{equation}
E(\mathbf{m})=P_{i}m_{i}+R_{ijk}m_{i}m_{j}m_{k}+...\,.\label{E_anom2}%
\end{equation}
The anomalous probability (\ref{p+}) then takes the form%
\begin{equation}
p^{+}(\mathbf{m})=\frac{1}{2}\left(  1+P_{i}m_{i}+R_{ijk}m_{i}m_{j}%
m_{k}+...\right)  \,,
\end{equation}
where, as noted, in general $P_{i}\neq(P_{i})_{\mathrm{QT}}$.

The term $R_{zzz}m_{z}m_{z}m_{z}$ corresponds to a correction%
\begin{equation}
\Delta p^{+}(\theta)=\frac{1}{2}R_{zzz}\cos^{3}\theta
\end{equation}
(again with the $z$-axis defined by $\mathbf{P}_{\mathrm{QT}}$). However, from
(\ref{Areas}), for $\theta=0$ we have $A_{-}=0$ while for $\theta=\pi$ we have
$A_{+}=0$. The total transmission probability $p^{+}(\theta)$ is therefore
subject to the constraints $p^{+}(0)=1$ and $p^{+}(\pi)=0$ for all finite
distributions $\rho$. To satisfy these constraints, we must include a
correction to the effective mean polarisation as well, with an anomalous
expectation value%
\begin{equation}
E(\theta)=(1-R_{zzz})\cos\theta+R_{zzz}\cos^{3}\theta\,.
\end{equation}
This corresponds to a corrected transmission probability%
\begin{equation}
p^{+}(\theta)=p_{\mathrm{QT}}^{+}(\theta)-\frac{1}{2}R_{zzz}\cos\theta
+\frac{1}{2}R_{zzz}\cos^{3}\theta\,.
\label{p_plus}%
\end{equation}
Note that the symmetry $E(-\mathbf{m})=-E(\mathbf{m})$ requires $E=0$ for
$\theta=\pi/2$, hence $p^{+}(\pi/2)=\frac{1}{2}$, which is consistent with (\ref{p_plus}).

For photon polarisation measurements along an angle $\Theta=\theta/2$ in
3-space, we may write the corrected transmission probability in the form%
\begin{equation}
p^{+}(\Theta)=p_{\mathrm{QT}}^{+}(\Theta)+\frac{1}{2}R_{zzz}f(\Theta)\,,
\end{equation}
with%
\begin{equation}
f(\Theta)=-\cos(2\Theta)\sin^{2}(2\Theta)
\label{f_theta}
\end{equation}
(where $f(\Theta)=0$ at $\Theta=0,\pi/4,\pi/2$). The usual Malus law
(\ref{Malus}) tested by Papaliolios then receives a correction%
\begin{equation}
\Delta p^{+}(\Theta)=\frac{1}{2}R_{zzz}f(\Theta)\,,
\label{delta_p_plus}
\end{equation}
where the magnitude of the coefficient $R_{zzz}$ is to be constrained by experiment.

\section{Testing short-timescale decay photons at high energies}
\label{sec:photons}
In this section, we consider short-lived systems whose decays produce high-energy photons, focusing on two prime examples: (i) top-quark decay, and (ii) neutral-pion decay. Both particles have extremely short lifetimes ($\sim10^{-25}\,\mathrm{s}$ for the top quark and $\sim10^{-17}\,\mathrm{s}$ for the neutral pion), making them interesting systems where one can put constraints on the Born rule at very short timescales.

\subsection{Top-quark decay}
The top quark, with a mass $m_t \approx 173\,\mathrm{GeV}$, has a decay width (decay rate)~\cite{EAP} 
\begin{equation}
  \Gamma_t \;\approx\; \frac{G_F\,m_t^{3}}{8\sqrt{2}\,\pi}
  \label{eq:top_width}
\end{equation}%
(where $G_F$ is the Fermi coupling constant), corresponding to an extremely short lifetime of $\sim 10^{-25}\,\mathrm{s}$. This timescale is so short that the top quark decays before hadronisation, meaning any photons emitted during top-quark decay emerge from a process with remarkably short timescales and large momentum transfers. In fact these are, in order of magnitude, the shortest timescales currently accessible at colliders, and therefore these photons serve as an ideal probe for testing the Born rule at very short times. 

\begin{figure}[H]
    \centering
    \subfigure[]{\includegraphics[width=0.48\textwidth]{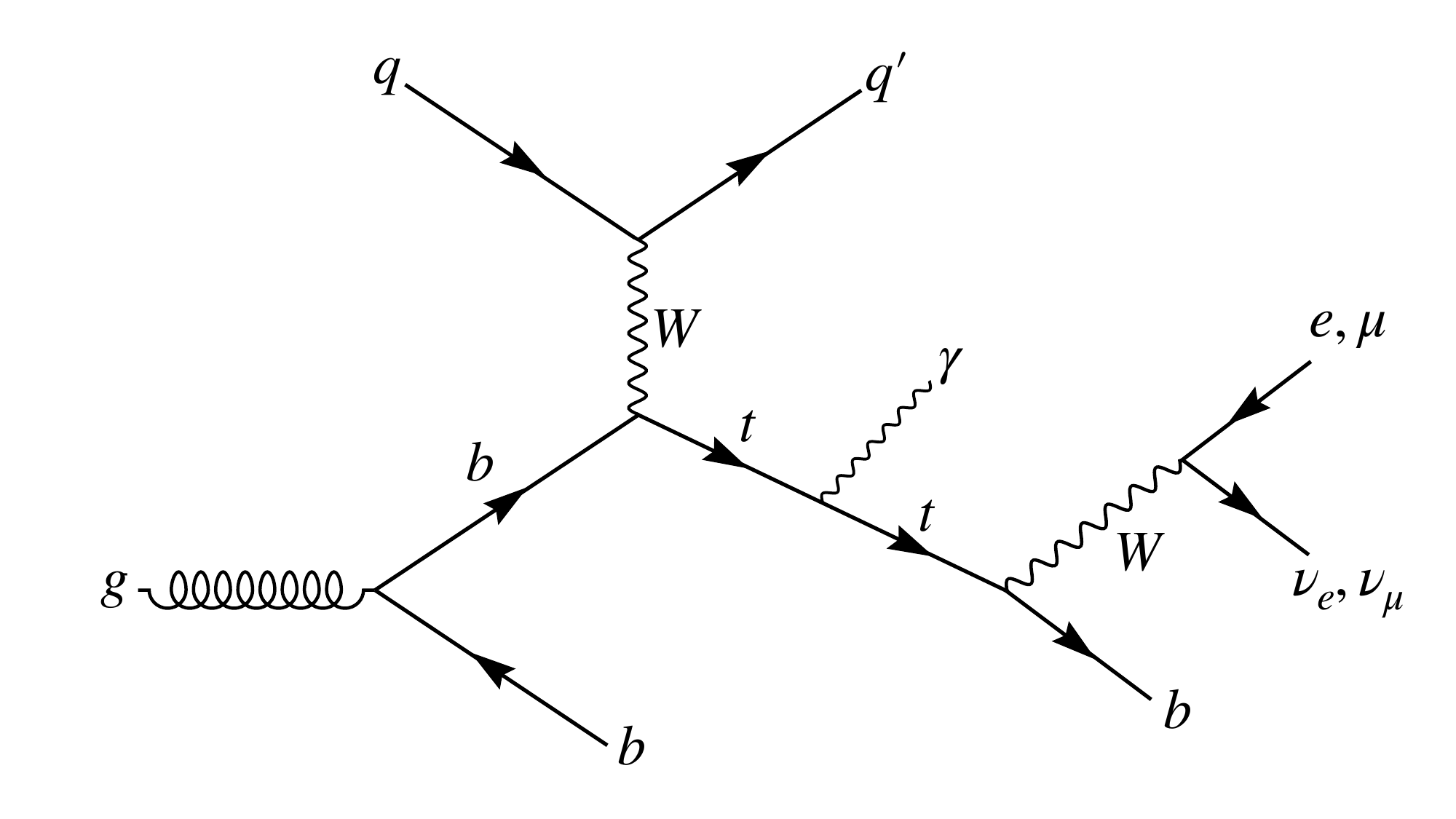}}
    \subfigure[]{\includegraphics[width=0.48\textwidth]{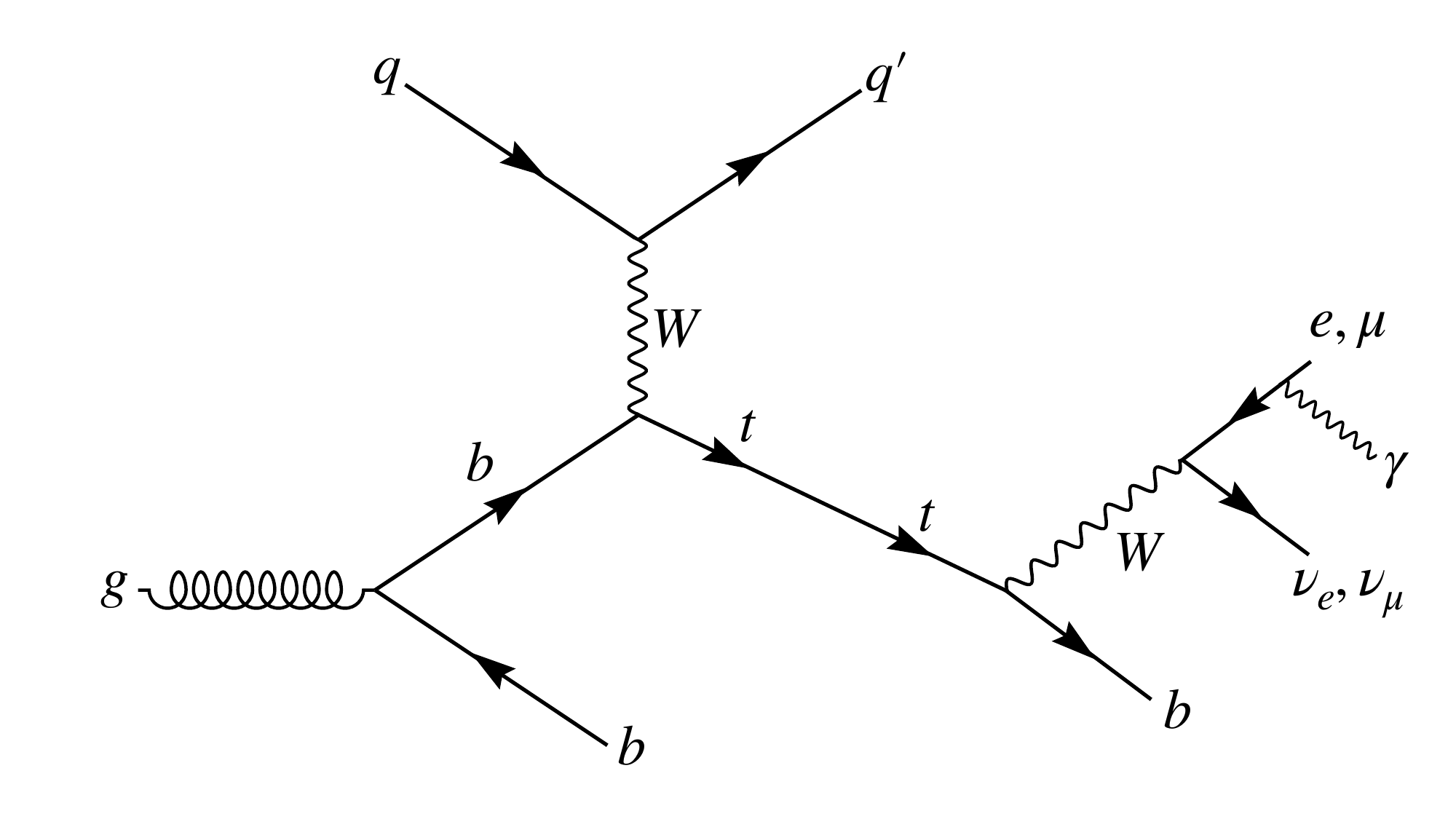}}
    \caption{Feynman diagrams showing (a) photon radiation from the top quark before decay and (b) photon radiation from a lepton in the decay products.}
    \label{fig:feynman}
\end{figure}

When detecting photons in top-quark events, one must distinguish between cases where the photon is radiated by the top quark before decay (Fig.~\ref{fig:feynman}a) or, alternatively, radiated by one of its charged decay products (Fig.~\ref{fig:feynman}b). Experimentally, this distinction is made by examining how the photon affects the mass reconstruction of both the top quark and the $W$-boson. If the photon was radiated by one of the decay products, including it in the mass reconstruction will improve our measurement of these masses. However, if the photon was radiated by the top quark itself, including it will not improve the mass reconstruction.

One might question why we specifically separate cases where the photon is radiated from the $W$-boson, given that the $W$-boson lifetime is also approximately $10^{-25}\,\mathrm{s}$, and so might just as well exhibit Born-rule violations. While photons originating from the $W$-boson or its decay products might also demonstrate comparable anomalous behaviour, no clear experimental distinction is made for these sources. Therefore, we focus on photons originating directly from the top quark, as this provides the cleanest sample for our analysis.

Recent measurements by the ATLAS Collaboration~\cite{ATLAS:2023qdu} have successfully demonstrated this technique, corresponding to a signal of $5{,}734$ photons from top-quark events with a total integrated luminosity of $139\,\mathrm{fb}^{-1}$, as inferred from the associated HEPData tables. The experiment distinguished between photons radiated directly from the top quark ($4{,}899$ events) and those from decay products ($835$ events) using the mass-reconstruction method described above.

To implement this approach, the ATLAS analysis relies on Monte Carlo (MC) simulations to estimate both the signal and background processes. A neural network (NN) is then trained on these MC-generated events, learning how to distinguish events with photons from single-top-quark production (where the photon is radiated either directly by the top quark or by a top decay product) from other background processes. This trained NN is subsequently applied to real data, using reconstructed kinematics (including mass-related observables) to classify events as signal-like or background-like.

\begin{figure}[H]
\centering
\adjustbox{trim={0 2cm 0 2cm},clip}{
\includegraphics[width=0.8\textwidth]{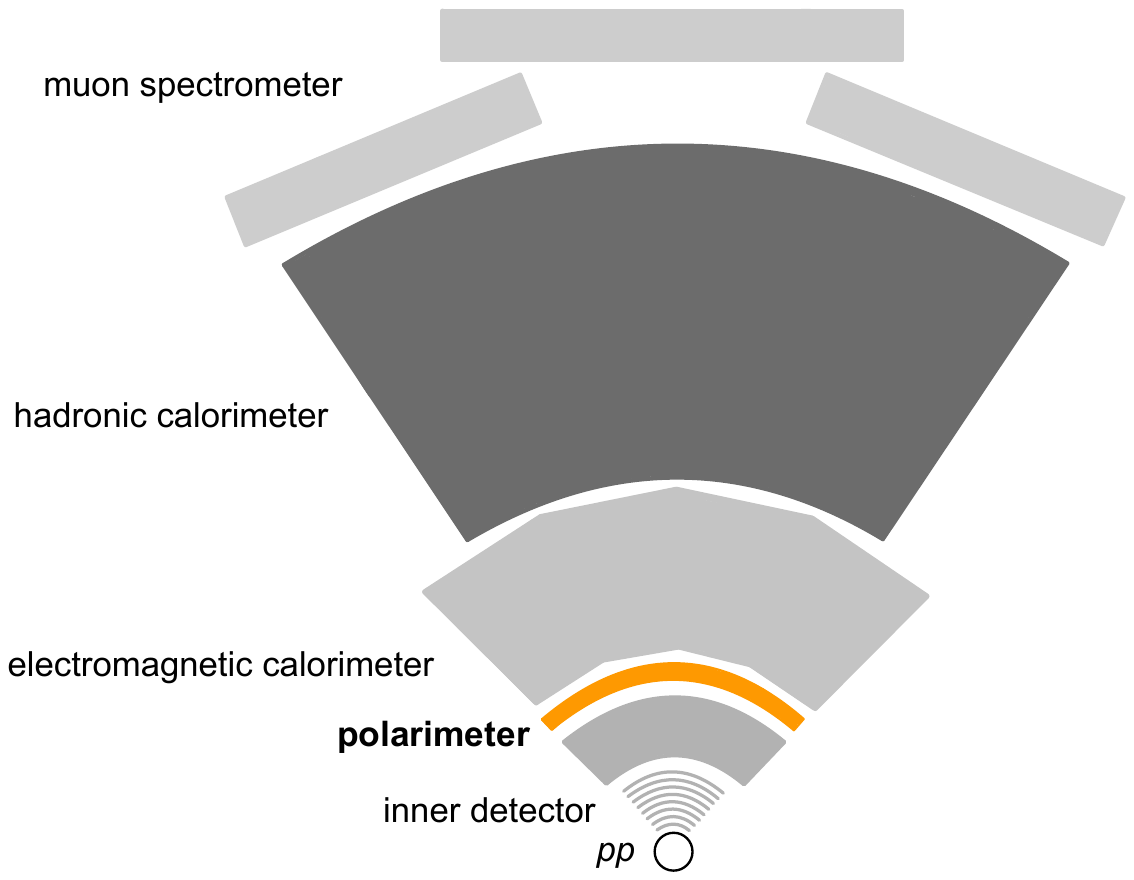}
}
\vspace{-1.4cm}
\caption{Schematic view of the ATLAS detector, showing the inner detector, electromagnetic calorimeter, hadronic calorimeter, and muon spectrometer. The proposed polarimeter (for individual or mean polarisation) would be integrated between the inner detector and EM calorimeter.}
\label{fig:detector}
\end{figure}

We propose to build upon this successful measurement by adding polarisation measurements, whether for individual photons or only ensemble averages. These measurements must be non-invasive, ensuring that all other parts of the ATLAS detector remain unchanged. The ATLAS detector~\cite{ATLAS:2008xda} identifies photons by combining information from multiple subsystems (Fig.~\ref{fig:detector}). First, the inner tracking detector ensures there is no charged-particle track associated with the candidate photon. Next, the electromagnetic (EM) calorimeter measures a localised electromagnetic shower. Finally, an isolation requirement is imposed: within a small angular cone around the photon candidate, the total energy measured in both the EM and hadronic calorimeters must remain below a specified threshold. This helps distinguish single photons from more energetic hadronic jets.

We propose adding a polarisation-measurement component (shown in orange in Fig.~\ref{fig:detector}) between the inner detector and the EM calorimeter, in a way that does not interfere with existing measurements. Since real data does not reveal the origin of a photon until its reconstructed kinematics are compared with MC predictions, all photons would undergo polarisation measurement. Only after the NN identifies each photon do we discard those labeled as decay-product emissions. In this way, only photons that are genuinely radiated by the top quark are kept, isolating events best suited for testing the Born rule.

If we want to probe the polarisation probability for individual photons, and study how it varies with the measurement axis, then it is important that we are able to choose and change the polarimeter measurement axis freely. In this way, a range of polarisation bases can be explored without modifying the rest of the detector. If instead we are measuring the average photon polarisation (where in this paper by average polarisation we mean the quantum theoretical average defined in (\ref{PQT})), a variable measurement axis is not necessary. Although integrating a polarimeter into the detector poses technical challenges, recent advances in high-energy photon polarimetry suggest feasible implementation strategies. Measuring the mean photon polarisation may be more straightforward in the near term; however, probing individual photon polarisations would allow one to study more subtle deviations from the Born rule.

\subsection{Neutral pion decay}
\begin{figure}[H]
\centering
\adjustbox{trim={0 0.5cm 0 0.5cm},clip}{
\includegraphics[width=0.55\textwidth]{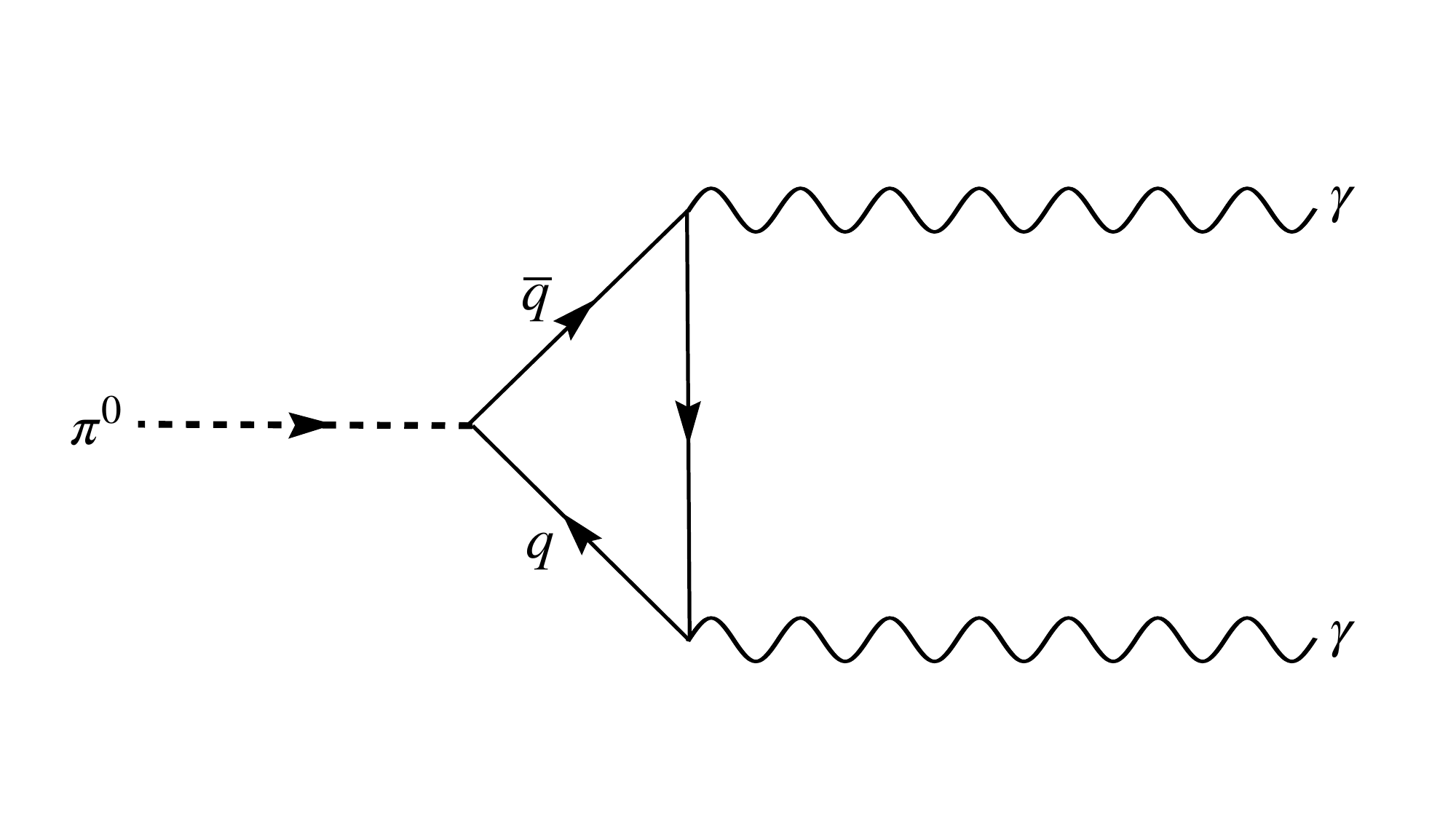}
}
\caption{Schematic diagram of a forward $\pi^0$ decay at the RHIC. A high-energy proton collides with another polarised proton (not shown), producing a forward $\pi^0$ that decays into two photons.}
\label{fig:pi0_feynman}
\end{figure}

Another useful example for testing the Born rule is neutral pion decay. The neutral pion, with a mass of approximately 135\,MeV~\cite{EAP}, decays into two photons with a lifetime of about $8.4 \times 10^{-17}\,\mathrm{s}$ (Fig.~\ref{fig:pi0_feynman}). Although this timescale is much longer than that of the top quark, it remains four orders of magnitude shorter than the timescale probed in the Papaliolios experiment~\cite{Pap67}. The $\pi^0$ system offers a crucial advantage: it decays almost exclusively into two photons (with a branching ratio of about $98.8\%$), minimising ambiguities in particle identification such as those associated with top-quark decay. 

Typically, $\pi^0$ mesons are produced in proton-proton ($pp$) collisions~\cite{RHIC-f:2020koe}. Since $pp$ collisions produce a variety of particles, one must separate the photons originating from $\pi^0$ decay ($\pi^0\rightarrow\gamma\gamma$) from those arising via other processes including accidental coincidences between photons from different $\pi^0$'s, direct photons, photons from $\eta$ decay, and misidentified neutrons. The procedure is analogous to that used for top quarks: if both photons originate from the same $\pi^0$, their invariant mass will reconstruct to $135\,\mathrm{MeV}$. If the two-photon mass falls outside a window of approximately $\pm3\sigma$ around this peak, the photon pair is classified as background. 

\begin{figure}[H]
\centering
\adjustbox{trim={0 0cm 0 1cm},clip}{
\includegraphics[width=0.8\textwidth]{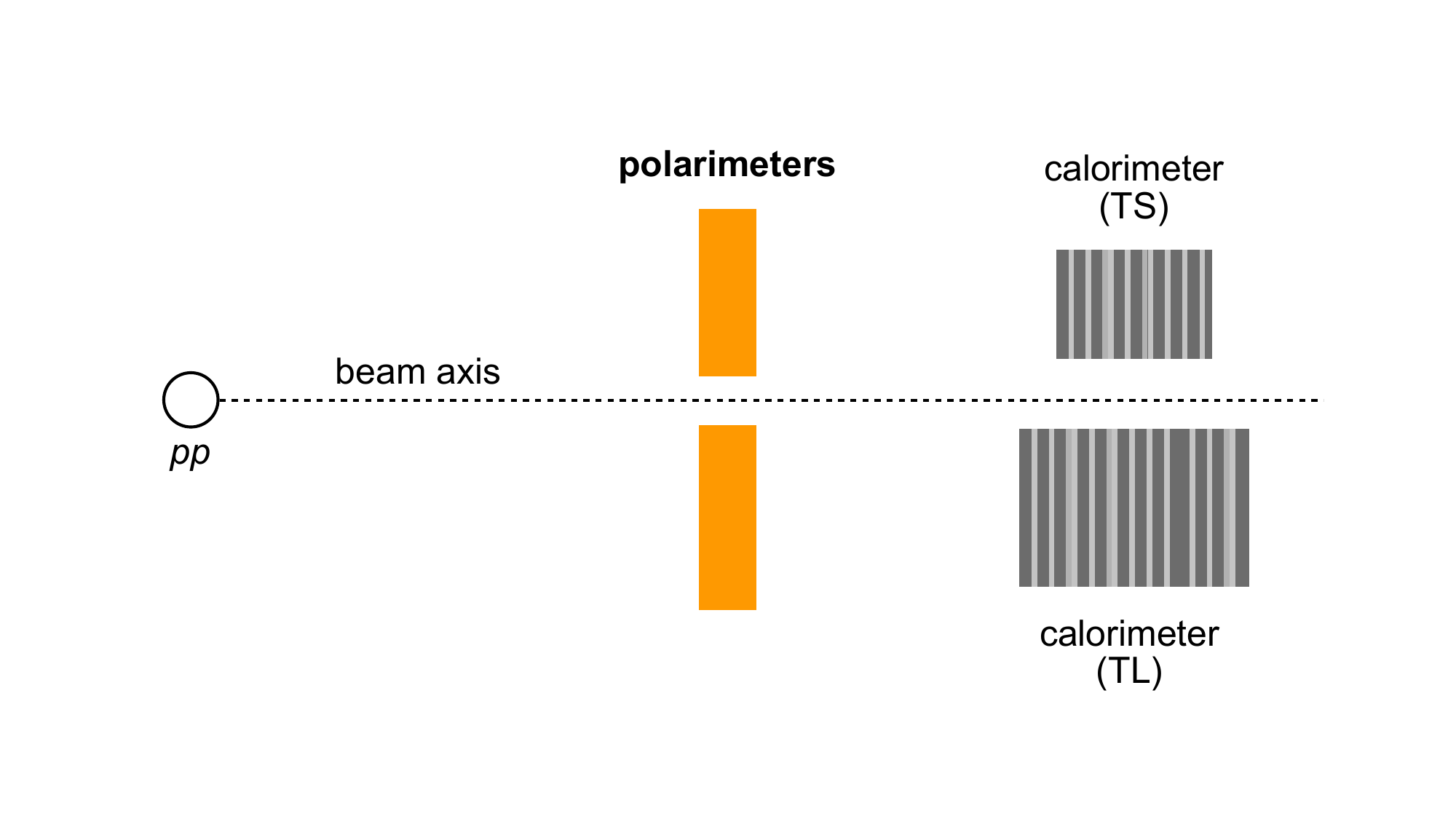}
}
\vspace{-1.1cm}
\caption{Schematic view of the RHICf detector, showing the $pp$ collision point (along the beam direction), the small tower, and the large tower. A non-invasive polarimeter could be placed before each tower.}
\label{fig:RHICF_detector}
\end{figure}

At the RHIC, for instance, the RHICf detector~\cite{RHICf:2021ahs} consists of two sampling calorimeters (a 20\,mm small tower and a 40\,mm large tower) positioned 18\,m from the collision point (Fig.~\ref{fig:RHICF_detector}). The detector can be moved to different vertical positions to cover a range of photon momenta. We propose adding a non-invasive polarisation measurement (for either individual or mean polarisation) in front of each calorimeter. In the case of individual photon measurements, the axis should be adjustable to measure different polarisation orientations. 

Events are classified based on how the $\pi^0$ decay photons interact with these towers. In Type-I events, the two photons are detected in separate towers, allowing clear separation and measurement of the energy and position of each photon. In Type-II events, however, both photons are detected within a single tower, requiring a more sophisticated method of analysis to disentangle the contribution of each photon. 

The RHICf Collaboration has demonstrated the feasibility of this measurement, collecting $1.1\times10^8$ events with an integrated luminosity of $\sim700\,\text{nb}^{-1}$. About $200,000$ decays $\pi^0\rightarrow\gamma\gamma$ were identified in the cleaner Type-I topology, with a background-to-signal ratio as low as $2\%$ at high momenta, providing an exceptionally clean sample for testing the Born rule.

\subsection{Experimental constraints and targets} 
\label{subsec:constraints_and_targets}

For mean polarisation measurements, for which there already exists technology widely used in astrophysics (Section~\ref{sec:introduction}), we might consider a small shift in the magnitude of the mean polarisation,
\begin{equation}
P = P_{\mathrm{QT}} + \varepsilon,
\end{equation}
where $P$, $P_{\mathrm{QT}}$ are respectively the magnitudes of $\mathbf{P}$, $\mathbf{P}_{\mathrm{QT}}$ and $\varepsilon$ is a small parameter. As noted in Section~\ref{sec:introduction}, however, current techniques allow us to measure mean photon polarisations at energies rather lower than those required at a collider. As will be discussed in Section~\ref{subsec:tau_lepton_bound}, mean polarisations have in fact already been measured for tau leptons, and for that case current experiments already allow us to put simple bounds on possible deviations from the quantum mean. Here, for the case of photons, we focus on a more ambitious long-term possibility, in which measurements of individual photon polarisations would allow a full probe of the Born rule. 

We now consider how accurately individual photon polarisations must be measured in order to place quantitative constraints on the Born rule. We provide bounds for: (i) the minimum number of events $N_{\min}$, (ii) the required transmission count accuracy $\Delta n$, and (iii) the allowed polariser angle misalignment $\Delta \Theta$. Our results provide what seem reasonable targets for the future development of single-photon gamma-ray polarimetry.

From Section~\ref{subsec:the_bell_model}, we know that for photon polarisation measurements along an angle $\Theta=\theta/2$, Malus’ law ($p_{\mathrm{QT}}^{+}(\Theta)=\cos^{2}\Theta$) potentially can receive a correction, which in its simplest form is given by (\ref{delta_p_plus}). Dropping the subscript $zzz$, we may write the model probability as
\begin{equation}
p^+(\Theta; R) = \cos^2 \Theta + \tfrac{1}{2} Rf(\Theta), 
 \label{model_prob}%
\end{equation}
where $f(\Theta)$ is given by (\ref{f_theta}). We wish to consider how to put quantitative experimental constraints on the anomalous parameter $R$, where of course in quantum theory $R = 0$.

At a fixed angular setting $\Theta_{\rm{set}}$, we consider the unbiased estimator for $R$, 
\begin{equation}
\hat R(\Theta_{\rm set})
\;=\;
\frac{2\big(\hat{p}^+-\cos^2\Theta_{\rm set}\big)}{f\big(\Theta_{\rm set}\big)},
 \label{estimator}%
\end{equation}
with 
\begin{equation}
\hat{p}^+ = \frac{n}{N},
\end{equation}
where $n$ is the number of transmitted photons out of $N$ trials.

\subsubsection{Minimum number of events $N_{\min}$} 

For fixed $\Theta_{\rm{set}}$, $\hat{p}^{+}$ is a binomial random variable with variance 
\begin{equation}
\sigma_{\hat{p}^{+}}^{2}=\frac{p^{+}(1-p^{+})}{N}. 
\end{equation}
By error propagation, the standard deviation for  $\hat{R}$ will be 
\begin{equation}
\sigma_{\hat R}
=\Big|\frac{d\hat R}{d\hat p^{+}}\Big|\sigma_{\hat p^{+}}
=\frac{2}{|f(\Theta_{\rm set})|}\sqrt{\frac{p^{+}(1-p^{+})}{N}}.
\end{equation}
In the large-$N$ (Gaussian) limit, the statistical significance associated with an observed value $\hat{R}$ is given by~\cite{Cowan:2010js} 
\begin{equation}
Z_0 = \frac{\hat R}{{\sigma_{\hat{R}}}}. 
\end{equation}
To claim a $3\sigma$ detection of a deviation $|R| = \delta$, so that $\sigma_{\hat R}\le \delta/3$, we then require 
\begin{equation}
N \geq \frac{36 p^+(1-p^+)}{\delta^2{[f(\Theta_{\rm set})}]^2}.
\end{equation}
Taking $p^+$ as (to a good approximation) the expected probability under the null hypothesis ($R = 0$, so $p^+=\cos^2\Theta$) gives 
\begin{equation}
N \geq \frac{36}{\delta^2\sin^2(4\Theta_{\rm set})}. \label{N_min}%
\end{equation}
At $\Theta_{\rm set} = \pi /8$, the right-hand side reaches its minimum so at other angles the minimum number of events will be larger. Taking, for illustration, $\delta = 10^{-2}$, this gives
\begin{equation}
N_{\min} \geq 3.6 \times 10^5,
\end{equation}
where $N_{\min}$ refers to the number of signal photons in the absence of background events. 
 
If instead the sample contains a mixture of signal ($N_s$) and background ($N_b$) events, we have a signal fraction $f_s = N_s/(N_s + N_b)$ and a total transmission probability 
\begin{equation}
p^+_{\rm tot}=f_s\,p^+_{s}+(1-f_s)\,p^+_{b},
\end{equation}
where $p^+_{s}$ and $p^+_{b}$ are respectively transmission probabilities for signal and background photons. If the background photons show no Born-rule deviation $(R_b = 0)$, then we find 
\begin{equation}
p^+_{\rm tot} = \cos^2\Theta_{\rm set} + \tfrac12 f_s R_{s} f(\Theta_{\rm set}),
\end{equation}
where $R_s$ is the anomalous parameter for the signal photons. This corresponds to an effective dilution of $R$, with $R_{\rm eff} = f_s R_s$. Since in (\ref{N_min}) we found  $N_{\rm min} \propto 1/\delta^2$, the required number of events in the presence of background becomes
\begin{equation}
N_{\rm min}' = \frac{1}{f_s^2} N_{\rm min}.
\label{N_min_with_bkg}
\end{equation}

For photons radiated by top quarks, the fraction of signal photons originating from the $tq\gamma$ process can be taken to be $f_s \simeq 0.8$~\cite{ATLAS:2023qdu}, since the $t(\to\ell\nu b\gamma)q$ contribution is found to be $\approx 20\%$ of the events in the fiducial region. Using (\ref{N_min_with_bkg}), this gives $N_{\rm min}' \approx 5.6 \times 10^5$. In the experiment, about $5 \times 10^3$ events were reported, which is two orders of magnitude smaller than our target $N_{\rm min}'$. For neutral pion decay, the backgrounds are negligible, and about $2 \times 10^5$ events were reported in the clean Type-I topology. This corresponds to about $4 \times 10^5$ photons, slightly exceeding our threshold $N_{\rm min} \geq 3.6 \times 10^5$.

\subsubsection{Transmission count uncertainty $\Delta n$}

A small shift in the transmission count $n \rightarrow n + \Delta n$ shifts $\hat{p}^+$ by 
\begin{equation}
\Delta \hat{p}^+ = \frac{\Delta n}{N}, 
\end{equation}
which leads to a shift in the estimator
\begin{equation}
\Delta \hat{R}_{\rm(n)}  = \frac{2 \Delta \hat{p}^+}{f(\Theta_{\rm set})}. 
\end{equation}
To prevent systematic effects from dominating the total uncertainty, we require 
\begin{equation}
|\Delta \hat{R}_{\rm(n)}| \leq \frac{\delta}{3}, 
\end{equation}
which corresponds to a bound of 
\begin{equation}
\Delta \hat{p}^+ \leq \frac{\delta}{6} |f(\Theta_{\rm set})|.
\end{equation}

Again taking $\Theta_{\rm set} = \pi / 8$, we have $|f(\Theta_{\rm set})| = 1/2\sqrt{2}$, hence
\begin{equation}
\Delta \hat{p}^+ \leq \frac{\delta}{12\sqrt{2}}, 
\end{equation}
which implies
\begin{equation}
\Delta n \leq \frac{\delta}{12\sqrt{2}} N. 
\end{equation}
As before, taking $\delta = 10^{-2}$ and with $N = N_{\rm min} = 3.6 \times 10^5$, we have $\Delta n \lesssim 212$. This corresponds to a fractional counting precision of 
\begin{equation}
\frac{\Delta n}{n} \approx 0.07\%
\label{eq:fractional_counting_precision} 
\end{equation}
(where $n$ is estimated to be $N\cos^2(\pi/8)$).  
 
\subsubsection{Angular misalignment $\Delta\Theta$} 
If the polarimeter is misaligned by a small angle $\Delta\Theta$, so that the true angle 
\begin{equation}
\Theta_{\rm true} = \Theta_{\rm set} + \Delta\Theta,
\end{equation}
with $|\Delta\Theta| \ll 1$, then the model probability (\ref{model_prob}) becomes $p^+(\Theta_{\rm true};R)$, and our estimator for $R$ will be 
\begin{equation}
\hat R
=\frac{2(p^+(\Theta_{\rm true};R)-\cos^2\Theta_{\rm set})}{f(\Theta_{\rm set})} =\frac{2\big(\cos^2\Theta_{\rm true}-\cos^2\Theta_{\rm set}\big)}{f(\Theta_{\rm set})}
+R\,\frac{f(\Theta_{\rm true})}{f(\Theta_{\rm set})}.
\end{equation}
Expanding to first order in $\Delta \Theta$, and neglecting the small $R$-dependent term, gives a shift in the estimator 
\begin{equation}
\Delta \hat{R}_{\rm(angle)} \simeq -2 \frac{ \sin(2\Theta_{\rm set})}{f(\Theta_{\rm set})} \Delta\Theta. 
\end{equation}

To ensure a statistically-limited measurement, and to claim a $3\sigma$ significance for a deviation $|R| = \delta$, we again require 
\begin{equation}
|\Delta \hat{R}_{\rm(angle)}| \leq \frac{\delta}{3}, 
\end{equation}
and so we have
\begin{equation}
|\Delta \Theta| \leq \frac{\delta}{6}\frac{|f(\Theta_{\rm set})|}{|\sin(2\Theta_{\rm set})|}.
\end{equation}
At $\Theta_{\rm set}= \pi/8$, and again for $\delta = 10^{-2}$, we find a maximum allowed angular misalignment
\begin{equation}
|\Delta \Theta| \approx 0.00083 \, \text{rad} \approx 0.05^\circ. \label{eq:delta_n_requirement} 
\end{equation}

\subsubsection{Benchmarks from optical polarimetry}

Optical photon polarimetry is a highly developed technology, and the accuracy required from our analysis is not unreasonable compared to what has already been demonstrated in quantum optics. A recent experiment using spontaneous parametric down conversion has measured polarisations for $1.7 \times 10^7$ individual photons~\cite{Branning:2010qzl}. The analysis showed that to keep polarisation bias below $1\%$, the half-wave plate must be aligned within $0.3^\circ$ of the ideal setting. This benchmark is only a factor of six less precise than our requirement (\ref{eq:delta_n_requirement}). In the same experiment, the average number of detections per time bin was measured to be $\mu = 0.364 \pm 0.002$. This corresponds to a fractional counting precision of $\Delta n/n = 0.002/0.364 \simeq 0.55\%$, about an order of magnitude less stringent than our target fractional counting precision (\ref{eq:fractional_counting_precision}). 

\section{Testing tau leptons with average helicity measurements}
\label{sec:tau_leptons}

Given that individual photon polarisations are not currently measurable at these energies, we now focus on an experiment that measures averages. One such approach involves measuring the average helicity of $\tau$ leptons produced in $Z$-boson decays (where helicity is, as usual, the spin along the axis defined by the momentum).\footnote{Here we must distinguish between average or mean polarisation as defined by (\ref{PQT}) (which makes no reference to any particular measurement axis), and the average helicity which is an average spin defined along a particular axis.} 

The $Z$-boson, with a lifetime on the order of $10^{-25}\,\mathrm{s}$~\cite{EAP}, again offers the possibility of probing the Born rule on the shortest timescales currently accessible. One must bear in mind, however, that we are not measuring the $\tau$ leptons directly at that extremely short timescale; rather, we observe their decay into $\pi^-\nu_{\tau}$ around $10^{-13}\,\mathrm{s}$ later. 

\begin{figure}[H]
\centering
\vspace{-0.2cm}
\adjustbox{trim={0 0cm 0 0cm},clip}{
\includegraphics[width=0.55\textwidth]{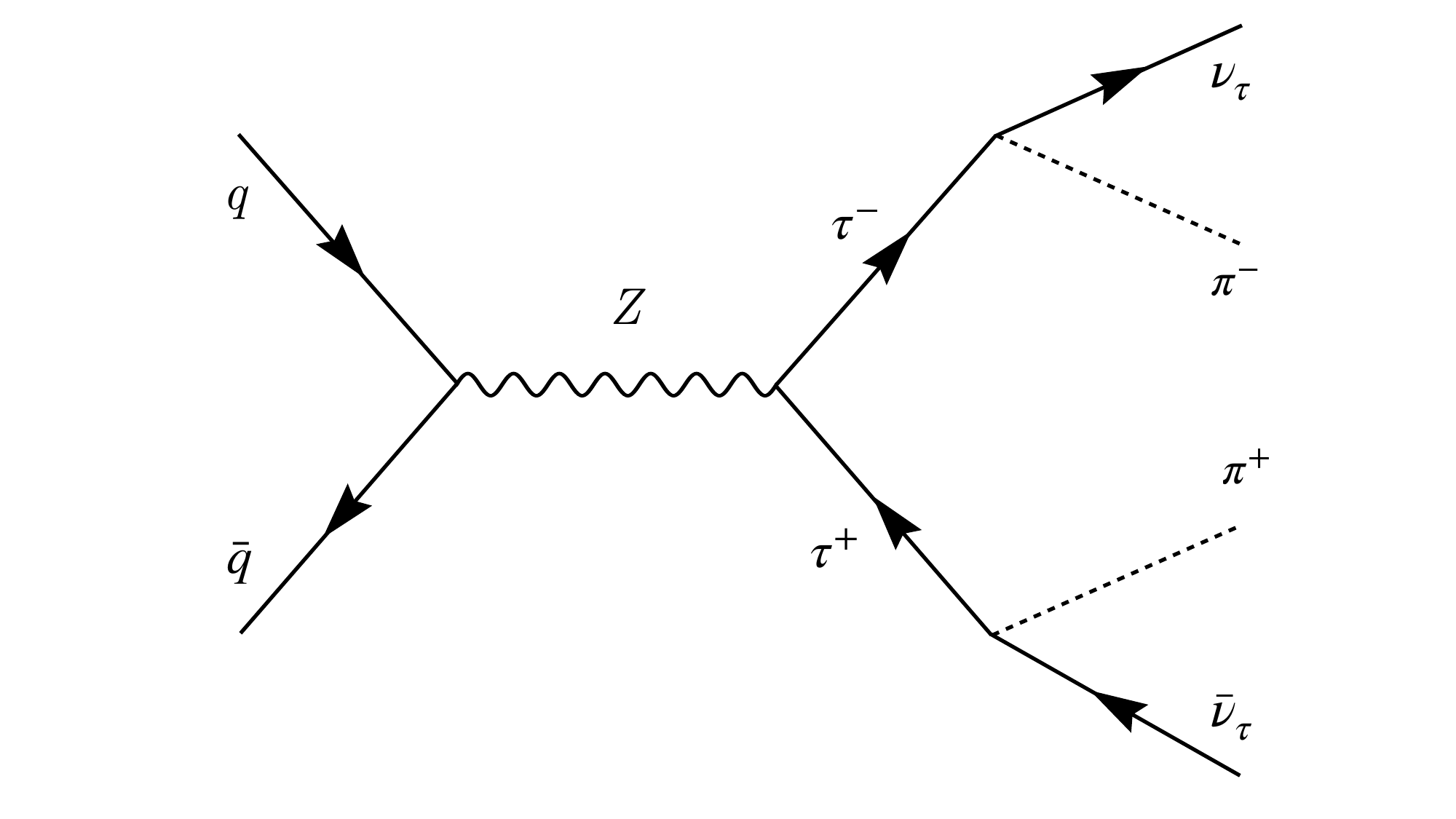}
}
\vspace{-0.1cm}
\caption{Feynman diagram of $Z$-boson production and decay to $\tau^+\tau^-$ in a $pp$ collision. Quarks from the colliding protons ($q$ and $\overline{q}$) annihilate to produce a $Z$-boson, which subsequently decays into a $\tau^+\tau^-$ pair. Each $\tau$ then decays into a tau neutrino ($\nu_{\tau}$ or $\overline{\nu}_{\tau}$) and a charged pion ($\pi^{-}$ or $\pi^{+}$).}
\label{fig:ZToTauTau}
\end{figure}

To produce $\tau$ leptons, $pp$ collisions at the LHC are typically used (see for example Fig.~\ref{fig:ZToTauTau} for a quark-antiquark annihilation into $Z$). Both CMS~\cite{CMS:2023mgq} and ATLAS~\cite{ATLAS:2017xuc} collect millions of $Z \rightarrow \tau \tau$ events with integrated luminosities above $100\,\mathrm{fb}^{-1}$. In this study, we focus on $\tau^+\tau^-$ pairs originating from $Z$ bosons. Because $pp$ collisions produce a wide range of final states, one must distinguish genuine $\tau$ leptons from other processes such as top-quark or $W$-boson decays. The procedure is analogous to how $\pi^0$ mesons are identified by their two-photon mass near $135\,\mathrm{MeV}$. Here, rather than looking for a $\gamma\gamma$ peak, we use specialised $\tau$-identification algorithms to reconstruct the $\tau^+\tau^-$ invariant mass near the $Z$-boson peak ($\sim91\,\mathrm{GeV}$). Events lying outside a suitable mass window (such as $\pm3\sigma$) are classified as background.

We may mention that a recent study has looked at reconstructing the complete density operator $\hat{\rho}$ for $\tau^{+}\tau^{-}$ pairs produced in $e^{+}e^{-}$ annihilation, and has used the resulting $\tau$‑polarisation observables to probe physics beyond the Standard Model (for example, setting limits on the $\tau$ electric dipole moment)~\cite{Fabbrichesi:2024xtq}. However, because this production channel does not originate from an especially short-timescale process, here we have instead focused on $\tau$ leptons originating from $Z$ decays. Also, instead of reconstructing the complete mean polarisation vector $\mathbf{P}$ (which for a quantum two-state system is equivalent to $\hat{\rho}$), we have limited ourselves to  reconstructing what is in effect the  longitudinal component of $\mathbf{P}$ (the mean helicity). Ensemble averages at colliders have also been explored for $t\bar{t}$ production at the LHC, where spin correlations in dileptonic decays may be used both to detect entanglement and to perform quantum tomography for the $t\bar{t}$ pair~\cite{Afik:2020onf}. More recently, polarisations reconstructed from scattering kinematics have been employed to test another fundamental feature of quantum theory, namely contextuality~\cite{Fabbrichesi:2025ifv,Fabbrichesi:2025rsg}. Some of that work looks at polarisation for $W$-bosons created by (very short-timescale) top-quark decays, and therefore may have some bearing on the physics explored here.

\subsection{Pure-state formula at fixed kinematics}

In a highly idealised situation, namely a single kinematic configuration at fixed partonic centre-of-mass energy $\sqrt{s}$ and fixed angle between the tau and its parent quark, $\theta$, the spin state of the $\tau^-\tau^+$ pairs can be written in amplitude form 
\begin{equation}
\label{eq:TauState}
|\tau\rangle = \; \mathcal{M}_{+}\,\lvert +-\rangle \;+\; \mathcal{M}_{-}\,\lvert -+\rangle,
\end{equation}
leading to a pure-state mean helicity (for the $\tau^{-}$) 
\begin{equation}
\label{eq:purestate_polarization}
\mathcal{P}_{\tau}^{\text{(pure)}}(s, \theta) = \frac{|\mathcal{M}_{+}|^2 - |\mathcal{M}_{-}|^2}{|\mathcal{M}_{+}|^2 + |\mathcal{M}_{-}|^2}.
\end{equation}
 
Let it be clear that $\lvert +-\rangle$ refers to $\tau^{-}$ with helicity $+\tfrac{1}{2}$ and $\tau^{+}$ with helicity $-\tfrac{1}{2}$, whereas $\lvert -+\rangle$ is the opposite. Thus, $\mathcal{M}_{+}$ corresponds to $\tau^{-}$ having helicity $+\tfrac{1}{2}$, and $\mathcal{M}_{-}$ corresponds to $\tau^{-}$ having helicity $-\tfrac{1}{2}$. No summation over angles or energies is included. In a perfect one-angle, one-energy scenario, such as an $e^+ e^-$ collision at a fixed $\sqrt{s}$ and fixed $\theta$, this amplitude-level expression is essentially the mean helicity of the sub-process.

\subsection{Ensemble-level cross-section formula}
\label{sec:ensemble_level_cross_section_formula}

Realistically, one integrates over multiple angles and possibly over a spread of parton energies. For two-body processes, the differential cross section is given by
\begin{equation}
\label{eq:twobody_crosssection}
\frac{d\sigma}{d\Omega}\bigg|_{\text{CM}} = \frac{1}{64\pi^2 s}\frac{p_f}{p_i}|\mathcal{M}|^2,
\end{equation}
where $d\Omega$ is the element of solid angle in the centre-of-mass frame. This integration over kinematic regions reflects the reality that at a collider it is practically impossible to produce all particles with identical energies and angles. Each collision with a certain final-state kinematics corresponds to a different pure state. 

However, because real data comprise many kinematic configurations (potentially across multiple runs), the ensemble is best described as a mixed state. Formally, one can write a density operator $\hat{\rho}$ that is a statistical mixture of these pure states
\begin{equation}
\label{eq:density_operator}
\hat{\rho} \;=\;\sum_{\text{kinematic bins}} \; w_i \;\lvert \tau_i\rangle \langle \tau_i\rvert,
\end{equation}
where each $\lvert \tau_i\rangle$ is of the form (\ref{eq:TauState}) but evaluated at different $s$ and $\theta$. The weights $w_i$ depend on the probability of obtaining a particular kinematic configuration and thus on the details of the scattering process. Consequently, the measured $\mathcal{P}_{\tau}$ at a collider is derived from a mixed state rather than a single pure state.

For every collision, each event corresponds to a pure state with different kinematics, each with its own $\mathcal{M}_{+}$ and $\mathcal{M}_{-}$. Since different runs have different kinematics, we have a mixed state. Once one integrates over all angles and parton energies, the ensemble-level (mean) helicity of the $\tau$ can be written in terms of helicity cross sections
\begin{equation}
\label{eq:ensemble_polarization}
\mathcal{P}_\tau^{\text{(ensemble)}} = \frac{\sigma_{+} - \sigma_{-}}{\sigma_{+} + \sigma_{-}},
\end{equation}
where $\sigma_+$ and $\sigma_-$ denote the cross sections for producing $\tau^-$ with helicity $+\tfrac{1}{2}$ and $-\tfrac{1}{2}$, respectively. In density operator language, each configuration corresponds to a pure spin state, whereas the final state is an overall mixture. Because hadron colliders typically sample many angles and energies, this cross-section ratio is the quantity we actually observe.\footnote{Many authors simply write $\mathcal{P}_{\tau}$ for the ratio (\ref{eq:ensemble_polarization}). Strictly speaking, it is still an average over fiducial phase space, so one could add angle brackets $\langle \cdots \rangle$. However, the shorthand without brackets is common in theory papers.}

\subsection{Extracting average tau helicity experimentally}
\label{sec:polarization_relation} 

Extracting the cross sections $\sigma_+$ and $\sigma_-$ requires precise information about the complete four-momenta of the $\tau$ leptons. However, because neutrinos are present in $\tau$ decays, direct measurement of their four-momenta is not experimentally feasible. Instead, experiments take advantage of the relation~\cite{Dick:2009}
\begin{equation}
\label{eq:event_distribution}
\frac{dN}{d\Omega} = I \frac{d\sigma}{d\Omega},
\end{equation}
where $I$ is the incident flux and $N$ is the observed number of events within a given kinematic configuration. Counting events in different angular bins thus serves as a proxy for the underlying cross section.

In the absence of detector effects and assuming full access to the $\tau$ rest frame,\footnote{In practice, $\tau$ rest frame kinematics are obtained probabilistically by reconstructing the missing neutrino momenta. Examples include the Missing Mass Calculator~\cite{Elagin:2010aw} (a likelihood scan over the kinematically allowed $\tau^+\tau^-$ phase space) and Point-Edge Transformer~\cite{Zhang:2025mmm} generative models that infer the neutrino momenta from event-level observables.} the two-body decay kinematics imply that the direction of the visible decay product is determined by the spin of the parent $\tau$, due to angular momentum conservation and the fixed spin states of the daughter particles. For $\tau^- \rightarrow \pi^- \nu_\tau$, the neutrino is always left-handed (with spin opposite to its momentum). Therefore
\begin{itemize}
\item A $\tau^-$ with spin $+\tfrac{1}{2}$ emits the $\pi^-$ along the spin direction ($\cos\theta > 0$);
\item A $\tau^-$ with spin $-\tfrac{1}{2}$ emits the $\pi^-$ opposite to the spin direction ($\cos\theta < 0$).
\end{itemize}
Here, $\theta$ is the angle between the $\tau$ spin direction and the pion momentum. Even though we are considering the $\tau$ in its rest frame and helicity is not strictly defined, the pion direction remains directly correlated with the spin state of the $\tau$. In this way, the pion emission angle serves as an experimental proxy for helicity.

We now define the forward region as $\cos\theta > 0$ and the backward region as $\cos\theta < 0$. Therefore, the experimentally observed mean helicity is defined as
\begin{equation}
\label{eq:forward_backward_asymmetry}
\langle \mathcal{P}_\tau \rangle_{\text{obs}} = \frac{N_f - N_b}{N_f + N_b}, 
\end{equation} 
where $N_f$ is the number of events in the forward region and $N_b$ is the number in the backward region. Because the helicity state is (ideally) in one-to-one correspondence with the emission direction of the pion, this ratio captures the asymmetry of the $\tau^-$ with respect to $+\tfrac{1}{2}$ and $-\tfrac{1}{2}$ helicity.

In the limit $N \gg 1$, assuming perfect detector efficiency, no background contamination, and precise knowledge of the $\tau$ rest frame and decay kinematics, we recover
\begin{equation}
\langle \mathcal{P}_\tau \rangle_{\text{obs}} = \mathcal{P}_\tau^{\text{(ensemble)}},
\end{equation}
justifying the use of the observed forward/backward decay counts as a proxy for the theoretical helicity asymmetry.

\subsection{Preliminary bound on deviations from the quantum mean} 
\label{subsec:tau_lepton_bound}

A microscopic model of the failure of the Born rule, in general, will give a correction to the mean polarisation. For tau leptons, this will imply an anomalous mean helicity
\begin{equation}
\langle P_\tau \rangle
= \langle P_\tau \rangle_{\mathrm{QT}} + \varepsilon,
\end{equation}
where $\langle P_\tau \rangle_{\mathrm{QT}}$ is the quantum-theoretical prediction and $\varepsilon$ is a small parameter. Existing ATLAS and CMS measurements of mean helicity for tau leptons, originating from decays $Z\!\to\tau^+\tau^-$, can then be interpreted as providing bounds on $\varepsilon$. 

For events where the partonic centre-of-mass energy $\sqrt{s}$ is known to be close to the $Z$-boson mass, we have 
\begin{equation}
\langle P_\tau \rangle_{\mathrm{QT}}= -\frac{2v_{\tau}a_{\tau}}{v^{2}_{\tau} + a^{2}_{\tau}}, 
\end{equation}
where $v_{\tau}$ and $a_{\tau}$ are the effective neutral-current and axial-vector couplings for the $\tau$ lepton~\cite{CMS:2023mgq}. For the case $v_{\tau} \ll a_{\tau}$, we may write  
\begin{equation}
\langle P_\tau \rangle_{\mathrm{QT}}
\approx -2 \frac{v_\tau}{a_{\tau}}= -2\left(1 - 4\,\sin^{2}\theta_{W}^{\rm eff}\right), 
\end{equation}
where $\theta_{W}^{\rm eff}$ is the effective weak mixing angle. Taking $\sin^{2}\theta_{W}^{\rm eff} = 0.2315 \pm  0.0002$~\cite{ALEPH:2005ab} we find
\begin{equation}
\langle P_\tau \rangle_{\mathrm{QT}} = -0.1472 \pm 0.0016. 
\end{equation}
This can be compared with the CMS fit result 
\begin{equation}
\langle P_\tau \rangle_{\mathrm{meas}} =-0.144 \pm 0.006 \ \rm{(stat)} \pm 0.014 \ \rm{(syst)}=-0.144\pm 0.015
\end{equation}
(where statistical and systematic uncertainties are reported separately and then combined).\footnote{This result includes multiple $\tau$ decay modes. If one were to solely focus on the $\tau^-\to\pi^-\nu_{\tau}$ decay channel, one would get cleaner results at the expense of less statistics.} 

If we interpret the experiment as it stands as providing a constraint on  $\varepsilon$, formally we find 
\begin{equation}
\varepsilon = 0.003 \pm \sigma_{\varepsilon}, 
\end{equation}
 where the total uncertainty $\sigma_{\varepsilon}$ is at least 0.015, since 
 \begin{equation}
 \sigma_{\varepsilon} = \sqrt{\sigma_{\text{meas}}^2 + \sigma_{\text{QT}}^2} \ge \sigma_{\text{meas}} = 0.015.
 \end{equation}
 Here, $\sigma_{\text{meas}}$ is the uncertainty from the measurement of mean helicity and $\sigma_{\text{QT}}$ is the uncertainty in the quantum-theoretical prediction coming from the uncertainty in the effective angle  $\theta_{W}^{\rm eff}$. If $\sigma_{\text{QT}}$ is neglected, we have a significance of only $0.2 \sigma$. Including the uncertainty in $\theta_{W}^{\rm eff}$ would reduce the significance further. The experiment therefore clearly favours $\varepsilon = 0$ and we can make no meaningful claim for a preferred $\varepsilon \neq 0$. 

To detect a deviation of $\varepsilon = 0.003$ at a $3\sigma$ level, we would need a total uncertainty of $\sigma_{\varepsilon} \approx 0.001$. The High-Luminosity LHC (HL-LHC) will have a luminosity increased by a factor of about $100$ compared to the cited CMS experiment, so the statistical uncertainty of $\langle P_\tau \rangle_{\mathrm{meas}}$ will be reduced by a factor of 10, from $\approx 0.006$ to about $0.0006$, which is already below our target. The limiting factor is therefore the systematic uncertainty, which in the cited CMS measurement is $\approx 0.014$.  

More than half of the systematic uncertainty ($0.008$ out of $0.014$) arises from decay-mode migration, in particular the uncertainty in the migration matrix that maps true hadronically decaying $\tau$ modes to their reconstructed decay channels. To reach $\sigma_{\varepsilon} \approx 0.001$ at the HL-LHC, we would need a $\tau$-related systematic uncertainty $\sigma_{\mathrm{syst}} \approx 10^{-3}$, corresponding to an improvement by a factor of about 20 compared with the current analysis ($\sigma_{\mathrm{syst}} = 0.014$). Achieving this reduction would require a significantly more accurate determination of $\tau$ decay modes, which is unlikely to be feasible without a focused experimental effort specifically optimised for $\tau$ polarisation.

\section{Background assumptions and Born-rule deviations}
\label{sec:bornrule_background}

It is important to note that in the experiments discussed in Sections~\ref{sec:photons} and~\ref{sec:tau_leptons}, the Born rule is assumed in the background modelling and event selection. Because the experimental procedure depends in part on the Born rule, we need to consider carefully how this might affect our ability to detect deviations from the Born rule. 

To illustrate how the experimental procedure depends on the Born rule, let us consider the example of background subtraction for mean tau helicity. The entire background-subtraction chain in this case relies on algorithms trained using MC samples generated under the Born rule. The process proceeds as follows. First, CMS (or another experiment) generates two dedicated MC samples: one where every $\tau$ is purely right-handed in helicity, and one where every $\tau$ is purely left-handed in helicity. These samples are simulated from the electroweak part of the Standard Model Lagrangian, where the corresponding Feynman amplitudes are extracted. For each kinematic point, the generator computes the amplitude $\mathcal{M}$, squares it to obtain $|\mathcal{M}|^2$, and accepts or rejects that point with probability proportional to $|\mathcal{M}|^2$. These simulated samples are then used to train an algorithm that differentiates between left-handed and right-handed helicity states in actual collision data. Similar issues arise in the photon cases (top-quark and $\pi^0$ decay). There, background modelling also depends on MC samples generated under the assumption that $d\sigma/d\Omega \propto |\mathcal{M}|^2$.

In principle, there is clearly a tension when searching for deviations from the Born rule with an experiment whose analysis, in part, depends on the Born rule. However, in this paper we are concerned with testing the Born rule only as applied to simple two-state systems (spin or polarisation) whereas the experimental modelling and event selection applies the Born rule to multi-particle scattering amplitudes. On this point, there are two points to make. First, it could happen that the Born rule is violated only in the first case and not in the second case. Second, because multi-particle scattering observables are averaged over broad regions of phase space, it seems plausible that small deviations from the Born rule could be washed out, leaving significant deviations only for the simple two-state systems. To address this point in future work, in principle we would need to model the entire experimental analysis, which does not seem feasible without some simplifying assumptions. Here we propose adopting a preliminary working hypothesis: we assume that the Born rule holds for background modelling and event selection, while allowing for possible deviations only in the two-state polarisation observables.

\section{Conclusion}
\label{sec:conclusion}

We have discussed how certain experiments at high-energy colliders might be repurposed as tests of the Born rule on timescales as short as $\sim 10^{-25}\,\mathrm{s}$. In ideal conditions, measuring the polarisation of individual gamma-ray photons would enable us to probe polarisation probabilities as a function of angle, providing detailed constraints on Born-rule violations, as parameterised here in terms of expectation values $E(\mathbf{m})$ that are nonlinear in the Bloch measurement axis $\mathbf{m}$. However, for such high-energy photons, current technology appears to permit measurement of average polarisation only, limiting the search for Born-rule violations to anomalies in the effective polarisation vector $\mathbf{P}$, which need not coincide with the usual quantum vector $\mathbf{P}_{\mathrm{QT}}$ as defined by (\ref{PQT}). We have illustrated this by measurements of mean helicity for tau leptons, enabling us to set a simple preliminary bound on Born-rule violations. 

Another concern relates to how the Born rule is employed in background modelling and event selection. One may broadly distinguish between potential Born-rule violations for spin or polarisation, and potential Born-rule violations for multi-particle scattering processes and final-state kinematics. We may reasonably assume that the latter would not necessarily disguise the former. Even so, this question deserves further study.

Quantum mechanics need not be taken for granted at high energies. Instead, high-energy data can and should be employed to place constraints on possible deviations from fundamental quantum principles, including the Born rule. Moreover, the resulting constraints can serve as a guide when attempting to construct theories beyond our current quantum physics. Pilot-wave theory, for example, requires regularisation at nodes of the wave function, and by this means can be extended to a model in which the Born rule is unstable at short timescales~\cite{AnnFond}. Experimental limits on Born-rule violations, at the shortest timescales currently accessible, will necessarily set tight constraints on such models.

It is also worth noting that several anomalies reported in the recent literature could potentially be related to Born-rule violations~\cite{HFLAV:2022esi}. If the Born rule is broken, then decay rates, and therefore ratios of decay rates, can deviate from Standard Model (SM) predictions. This makes some of the anomalies particularly intriguing. One example concerns anomalies in ratios of the $B$-meson decay widths in different decay channels, which collectively appear to show a deviation of about $3.3\sigma$ from the SM expectations. A related anomaly appears in $b \rightarrow s \ell^+ \ell^-$ transitions, where certain branching fractions for $B$-meson decays have been found to be up to $3.6\sigma$ lower than the SM prediction~\cite{LHCb:2021zwz}. Additionally, in the decay $B \rightarrow K^* \mu^+ \mu^-$, discrepancies have been observed in CP-averaged angular observables~\cite{LHCb20lmf,CMS24atz}, which are sensitive to interference between helicity amplitudes and may therefore act as probes of Born-rule violations. Finally, anomalies at the $2-3\sigma$ level have been reported~\cite{PDG:2024cfk} in the first row of the Cabibbo-Kobayashi-Maskawa (CKM) matrix, whose entries measure amplitudes for quark flavour transitions, and whose apparent unitarity violation could perhaps be related to some of the ideas discussed in this work.  

\textbf{Acknowledgements.} We warmly thank Oliver K. Baker for his interest in and support of this work. MV is also grateful to Paul L. Tipton for his helpful advice while revising this paper. MV is supported by the DOE Office of High Energy Physics under Grant No. DE-SC0017660.

\end{document}